\documentclass[conference,compsoc]{IEEEtran}

\usepackage[utf8]{inputenc} % allow utf-8 input
\usepackage[T1]{fontenc}    % use 8-bit T1 fonts
\usepackage{url}            % simple URL typesetting
\usepackage{booktabs}       % professional-quality tables
\usepackage{nicefrac}       % compact symbols for 1/2, etc.
\usepackage{microtype}      % microtypography

\usepackage{makecell}
\usepackage{multirow}
\usepackage{mathtools}
\usepackage{scalerel}
\usepackage{stackengine}
\DeclarePairedDelimiter\abs{\lvert}{\rvert}

\usepackage{amsmath,amsfonts}

\usepackage{amssymb}
\usepackage{amsthm}
\usepackage{placeins}
\usepackage[ruled,vlined,linesnumbered]{algorithm2e}
\SetKwProg{Fn}{Function}{:}{}

\makeatletter
\renewcommand\paragraph{\@startsection{paragraph}{4}{\z@}%
  {0.7ex plus 0.2ex minus 0.1ex}%
  {-0.6em}%
  {\normalfont\bfseries}}
\makeatother

\usepackage[dvipsnames]{xcolor}
\usepackage{graphicx}

\usepackage{fontawesome}
\usepackage{colortbl}
\usepackage{enumitem}
\usepackage{siunitx}

\usepackage{threeparttable}
\usepackage{tabularx}
\newcolumntype{L}{>{\raggedright\arraybackslash}X}
\newcolumntype{R}{>{\raggedleft\arraybackslash}X}
\newcolumntype{C}{>{\centering\arraybackslash}X}

\usepackage[numbers,sort&compress]{natbib}
\renewcommand{\citep}[1]{\cite{#1}}
\renewcommand{\citet}[1]{\cite{#1}}

\title{O2N2: Efficient Doubly Oblivious Approximate Nearest Neighbor Search}
\author{%
\IEEEauthorblockN{Zhechen Li\IEEEauthorrefmark{1},
Zimai Guo\IEEEauthorrefmark{2},
Mingxun Zhou\IEEEauthorrefmark{3},
Chang Shi\IEEEauthorrefmark{2},
Cheng Hong\IEEEauthorrefmark{1}}
\IEEEauthorblockA{\IEEEauthorrefmark{1}Ant Group \qquad
\IEEEauthorrefmark{2}Peking University\\
\IEEEauthorrefmark{3}The Hong Kong University of Science and Technology}
\IEEEauthorblockA{\texttt{lizhechen@pku.edu.cn},
\texttt{endergzm@pku.edu.cn},
\texttt{mingxunz@ust.hk}\\
\texttt{2300017795@stu.pku.edu.cn},
\texttt{vince.hc@antgroup.com}}
}

\newtheorem{theorem}{Theorem}
\newtheorem{definition}{Definition}
\newtheorem{proposition}{Proposition}
\newtheorem{lemma}{Lemma}

\theoremstyle{definition}
\newtheorem{remark}{Remark}
\newtheorem{claim}{Claim}

\def\bigo{ \operatorname{O} }

\def\dummy{\perp}

\def\prob{\mathbb{P}}
\def\N{\mathbb{N}}
\def\R{\mathbb{R}}

\def\E{\mathbb{E}}

\def\negl{\operatorname{negl}}
\def\dist{\operatorname{dist}}
\def\nn{\operatorname{NN}}
\def\view{\mathsf{Access}}

\def\osort{\mathsf{OSort}}

\def\pos{\mathsf{pos}}
\def\ptr{\mathsf{ptr}}
\def\lptr{\mathsf{lptr}}
\def\rptr{\mathsf{rptr}}
\def\root{\mathsf{root}}
\def\idx{\mathsf{idx}}

\def\bb{\bf\boldmath}

\newcommand{\kdt}{$k$-\lowercase{d} tree}
\newcommand{\name}{O2N2}
  \newcommand\dddagger{%
  \sbox0{\ddag}\stretchrel*{%
  \stackengine{-.78\ht0}{\ddag}{\ddag}{O}{c}{F}{F}{S}}{\ddag}%
    }

\begin{document}

\maketitle

\begin{abstract}
\boldmath
  Privacy-preserving approximate nearest neighbor (ANN) search allows clients to search from an untrusted server without revealing sensitive queries, and is useful in  scenarios like medical diagnostics and credit scoring. Among the various privacy-preserving primitives, trusted execution environments (TEEs) have much lower overhead than heavy cryptographic ones, yet they leak memory access patterns, enabling adversaries to infer query contents.

  Oblivious RAM (ORAM) can be used to hide those access patterns at $\bigo(\log N)$ asymptotic cost, but existing high-performance ANN search algorithms (e.g., IVF or HNSW) exhibit complex, data-dependent access patterns that are unfriendly for ORAM.  Furthermore, standard ORAM constructions typically assume there exists a small amount of trusted storage, which weakens the security guarantees and is not always feasible. Solutions that eliminate such trusted storage (known as {\em double obliviousness}) typically degrade performance further to $\bigo(\log^2 N)$.

  We introduce \name{}, a doubly oblivious ANN framework built around single-path search in randomly rotated \kdt{} forests. A pointer-based Circuit ORAM stores the search structure, with separate storage for compact internal nodes and vector-bearing leaves. Each visited leaf is processed by an oblivious selection network, and fixed-schedule duplicate removal and selection combine the local results into the final top-$k$.

  We evaluate \name{} on real-world datasets and further test its scalability on 100 million vectors. For 10-NN and 100-NN search at the specified high-recall targets, it achieves a $9.8$--$125.9\times$ speedup over the fastest doubly oblivious baselines. It also achieves comparable efficiency against existing singly oblivious systems such as Compass (OSDI~'25) and BiSON (EuroS\&P~'26), while providing stronger privacy guarantees.
\end{abstract}

\section{Introduction}
\label{sec:introduction}

\begin{table*}[t]
    \centering
    \caption{Comparison of private ANN search solutions, where {\em single (external) obliviousness} hides storage access patterns, while {\em double obliviousness} also hides access patterns inside the TEE.}
    \label{tab:comparison}
    \setlength{\tabcolsep}{4pt}
    \begin{tabularx}{\linewidth}{lcc>{\centering\arraybackslash}X>{\centering\arraybackslash}X}
        \toprule
        \textbf{Work} & \textbf{Data model} & \textbf{Core primitive} & \textbf{Privacy guarantee} & \textbf{Underlying ANN scheme} \\
        \midrule
        Panther~\cite{li2025panther} & Server-owned & PIR & Query privacy + database privacy & Cluster-based \\
        Pacmann~\cite{zhou2024pacmann} & Public & PIR & Query privacy & Graph-based \\
        Tiptoe~\cite{henzinger2023private} & Public & PIR & Query privacy & Cluster-based \\
        \midrule
        Compass~\cite{zhu2025compass} & Client-owned & ORAM & \makecell{Query privacy \\
        Single (external) obliviousness} & Graph-based \\
        BiSON~\cite{das2026bison} & Client-owned & ORAM & \makecell{Query privacy + database privacy \\
        Single (external) obliviousness} & Graph-based \\
        Onyx~\cite{rathee2026onyx} & Client-owned & ORAM & \makecell{Query privacy + database privacy \\
        Single (external) obliviousness} & Graph-based \\
        \rowcolor{gray!20}
        \textbf{\name{} (Ours)} & \textbf{Any} & \textbf{ORAM} & \makecell{Query privacy + database privacy \\
        \textbf{Double obliviousness}} & {\bf Tree-based} \\
        \bottomrule
    \end{tabularx}
\end{table*}

Approximate nearest neighbor (ANN) search is a core primitive for modern data-intensive applications, including embedding search and retrieval-augmented generation~\cite{mikolov2013distributed,lewis2020retrieval}. Given a query vector, ANN search retrieves its (approximately) top-$k$ closest vectors from a vector database, typically using specialized indexes such as IVF~\cite{jegou2010product}, HNSW~\cite{johnson2019billion}, and DiskANN~\cite{jayaram2019diskann} to avoid exhaustive scans. With the growth of database size, ANN search relies more and more on external memory, which is often untrusted. This creates a need for private ANN search schemes that protect both sensitive queries and database contents. Trusted Execution Environments (TEEs)~\citep{mckeen2013innovative,kaplan2016amd,zhu2024confidential} provide an attractive foundation for this goal by offering low-overhead protected execution on untrusted platforms~\cite{sardar2023confidential,feng2024survey}, where the enclave plays the role of trusted client in conventional outsourced storage systems.

% with real-world deployments such as Apple's Private Cloud Compute~\cite{ApplePCC2024} and Amazon Data Clean Rooms~\cite{AWSCleanRooms}.

However, TEEs alone do not solve the private ANN search problem. Although TEEs protect the confidentiality of code and data, they do not hide memory access traces. Prior attacks show that such traces can leak sensitive information~\cite{cash2015leakage,grubbs2016breaking,islam2012access,zhang2016all}. A standard defense against these attacks is to achieve {\em obliviousness}~\citep{goldreich1987towards}, which requires the access patterns to be data-independent. Traditional schemes treat the client-side storage as trusted and do not hide accesses to it. In TEEs, however, the enclave plays this client's role, and its internal accesses may be exposed through page-fault or cache-timing attacks~\cite{van2017telling,xu2015controlled,tinoco2023enigmap}. We therefore require \emph{double obliviousness}~\cite{mishra2018oblix,zheng2025h2o2ram,tinoco2023enigmap}, which requires both external and enclave memory access patterns to be independent of the sensitive input.

To achieve this guarantee, a natural black-box approach is running a non-private ANN index under a fixed, padded schedule, with all memory accesses routed through a doubly oblivious memory layer, such as doubly oblivious RAMs~\citep{zheng2025h2o2ram,mishra2018oblix} or doubly oblivious maps~\citep{mishra2018oblix,tinoco2023enigmap,chamani2023graphos}. This black-box route mitigates the access pattern leakage, but most of existing constructions typically pay $\Omega(\log^2 N)$ cost per logical access. Since mainstream ANN indexes perform many fine-grained, adaptive probes, this per-access overhead is repeatedly incurred and leads to impractical latency in our evaluation. This gap raises the following question.

\begin{center}
    \em How can we design a doubly-oblivious ANN search scheme that achieves both strong privacy guarantees and practical efficiency?
\end{center}

We answer this question with \name{}, a doubly oblivious ANN search system built around structural alignment between the search algorithm and the oblivious primitive. Defeatist search in a randomly rotated \kdt{} forest follows one root-to-leaf path per tree~\cite{ram2019revisiting}. \name{} combines pointer-based Circuit ORAM access~\cite{wang2015circuit,wang2014oblivious}, a heterogeneous layout for internal and leaf nodes, and oblivious selection networks over the complete leaf candidates. Independent tree queries expose multiple candidate sets, while the selection stage retains their best distinct neighbors.

Using the combined network by Cong et al.~\citep{cong2025revisiting}, \name{} answers 10-NN queries on SIFT1M at $90.01\%$ recall in $1.37$~ms. We evaluate latency, recall, and distance error across workloads, and use a paired ablation to isolate the effect of the candidate selector.

\subsection{Our contributions}

We summarize the main guarantee of \name{} below.

\begin{theorem}[Main result, informal]
    \label{thm:main-informal}
    For a dataset of $N$ points in $\R^d$ and a query for top-$k$ neighbors, \name{} provides a doubly oblivious ANN search scheme using a forest of $m$ trees, achieving a worst-case time complexity of $\bigo\bigl(m\log^2 N+mn_0d\log N+mk\log^2(mk)\bigr)$, where $n_0$ is the size of \kdt{} leaves.
\end{theorem}

For fixed $m$ and $k$, setting $n_0=\Theta(\log N)$ gives a query time of $\bigo(d\log^2 N)$. The forest size $m$ controls the tradeoff between query cost and recall, and its dependence is explicitly retained in our complexity bound.

Our contributions are summarized as follows:

\begin{itemize}[leftmargin=*]
    \item We formulate the problem of doubly oblivious ANN search and design \name{}, a practical construction that aligns randomized \kdt{} defeatist search with tree-based ORAM. This design avoids black-box doubly oblivious compilation of adaptive ANN probes and achieves logarithmic structural overhead relative to the underlying root-to-leaf search pattern.

    \item We design and implement two doubly oblivious ANN baselines, OIVF and OHNSW, by adapting IVF and HNSW to doubly oblivious storage and replacing data-dependent components with oblivious operations. These constructions may be of independent interest for secure vector search, with OHNSW achieving latency comparable to Compass-in-TEE~\citep{rathee2026onyx} while also protecting internal memory accesses.
\end{itemize}
Our code base is open-sourced~\footnote{\url{https://anonymous.4open.science/r/O2N2-B448}}.

\section{Related work and discussions}
\label{sec:related}

\paragraph{Private ANN with cryptographic primitives}
Early private ANN systems often relied on fully homomorphic encryption~\cite{zuber2021efficient,shaul2020secure}, which protects computation but tends to require expensive distance evaluation over large candidate sets. Another line of work improves efficiency under a two-server non-collusion assumption~\cite{servan2022private,song2023secure}, trading performance for an additional trust assumption. In the single-server setting, SANNS~\cite{chen2020sanns} and Panther~\cite{li2025panther} target stronger data-privacy settings using cryptographic primitives such as PIR, while Tiptoe~\cite{henzinger2023private} and Pacmann~\cite{zhou2024pacmann} assume a public database and focus on query privacy. Table~\ref{tab:comparison} summarizes these differences.

\paragraph{TEE- and ORAM-based private ANN}
ORAM is a natural complement to TEEs because it hides which encrypted records are accessed while allowing the computation itself to run efficiently inside protected execution. The closest system to ours is Compass~\cite{zhu2025compass}, which adapts HNSW~\cite{johnson2019billion} using Ring ORAM~\cite{ren2015constants}. Compass provides single external obliviousness and is therefore effective against an adversary observing outsourced storage, but it does not protect all enclave-internal access behavior. \name{} instead targets double obliviousness, which is needed when the adversary can observe both external memory and internal memory-access side channels.

\paragraph{Doubly oblivious RAM and black-box compilation}
Doubly oblivious RAM removes the trusted-storage assumption by making all memory accesses, including the accesses to ORAM metadata, oblivious~\cite{mishra2018oblix,zheng2025h2o2ram}. This stronger guarantee is expensive when used as a black-box compiler: every logical access of an ANN index incurs doubly oblivious memory overhead, while modern non-private ANN methods such as IVF~\cite{jegou2010product}, HNSW~\cite{johnson2019billion}, and DiskANN~\cite{jayaram2019diskann} rely on adaptive, data-dependent probing. \name{} follows a different route: it chooses a tree-based ANN algorithm whose query path can be made to coincide with a tree-based ORAM access pattern, reducing the mismatch between efficiency and obliviousness.

\section{Preliminaries}
\label{sec: prelim}

In this section, we formally define the problem of approximate nearest neighbor search and introduce the key cryptographic and algorithmic primitives that form the basis of our scheme.

\subsection{Approximate nearest neighbor search}

Let $\mathcal{X}$ be a metric space, where the distance between $p,q\in\mathcal{X}$ is denoted by $\dist(p,q)$. Let $D = \{p_1,p_2,\ldots, p_N\} \subset \mathcal{X}$ be a dataset of $N$ points in $\mathcal{X}$. Given $k\in\mathbb{N}_+$, the {\em $k$-nearest neighbor ($k$-NN for short)} search problem is defined as follows.

\begin{definition}[$k$-NN search]
    \label{def: kNN}
    Given an integer $k\in\N_+$, a dataset $D \subset \mathcal{X}$, and a query point $q \in \mathcal{X}$, the (exact) $k$-NN search problem is to find a point set $\mathcal{A}\subset D$ containing exactly $k$ points, satisfying that for any $p\in \mathcal{A}$ and $p'\in D\backslash \mathcal{A}$, $\dist(p,q) \leq \dist(p', q)$. The set $\mathcal{A}$ is called the {\em $k$-NN} of $q$ in $D$, denoted by $\nn(k,q,D)$.
\end{definition}

While the $k$-NN can be defined for any metric space, in the paper, we only focus on the $d$-dimensional Euclidean space $\R^d$. The distance between any two points $p, q \in \mathbb{R}^d$ is $\dist(p, q) = \|p - q\|_2$. Accordingly, the $k$-NN of point $q$ in $D$ is defined as
\begin{align*}
\nn(k,q,D) = \arg\min\nolimits_{A\subset D, |A|=k} \sum\nolimits_{p\in A} \dist(p,q).
\end{align*}

% For low-dimensional cases, e.g., when $d=\bigo(1)$, several tree-based algorithms can solve the $k$-NN search problem efficiently, such as \kdt{}~\citep{bentley1975multidimensional}, random projection tree~\citep{kleinberg1997two}, cover tree~\citep{beygelzimer2006cover}, and vantage point tree~\citep{yianilos1993data}. However, with the growth of dimension, especially when $d=\omu(\log n)$, the efficiency of these tree-based methods is severely hampered by the curse of dimensionality. For example, \kdt{} has a worst time complexity of $\bigo(n^{1-1/d})$~\citep{lee1977worst}, which rapidly approaches a $\th(n)$ linear scan as $d$ increases. In fact, to the best of our knowledge, there has not been any algorithm that can solve the (exact) $k$-NN search problem with $\bigo((\log n)^{\bigo(1)})$ worst case time complexity, which indicates that finding the true $k$-NN is computationally expensive in high-dimensional spaces. Consequently, many practical applications turn to approximate solutions, i.e., to find ANNs. An ANN search algorithm can be defined as a mapping $f$, which maps each tuple $(k,q,D)$ to a set containing $k$ points in $D$. The effectiveness of ANN search algorithms is usually measured by recall and distance ratio. Specifically, letting $f$ be an ANN search algorithm,

In practice, exact $k$-NN search is often too expensive for large, high-dimensional datasets, so practical systems typically use approximate nearest neighbor (ANN) search. We model an ANN search algorithm as a possibly randomized mapping $f$ that maps each tuple $(k,q,D)$ to a set of $k$ points in $D$. We measure its effectiveness using the following metrics.

\begin{itemize}[leftmargin=*]
    \item {\em Recall} measures the expected fraction of the true $k$-NN retrieved by the algorithm. Formally, given dataset $D$ and a probability distribution $Q$ of queries, the recall of $f$ is
    \begin{align*}
        \text{Recall@}k = \E_{q\sim Q} [|f(k,q,D)\cap \nn(k,q,D)|] / k.
        % \text{Recall@}k = \frac{1}{k}\cdot \E_{q\sim Q} [|f(k,q,D)\cap \nn(k,q,D)|].
    \end{align*}
    \item {\em Distance-sum ratio (DSR)} is the ratio of the distance sums between the approximate $k$-NN and the true $k$-NN, i.e.,
    \begin{align*}
        \text{DSR@}k = \frac{\sum\nolimits_{p\in f(k,q,D)} \dist(p, q)}{\sum\nolimits_{p\in \nn(k,q,D)} \dist(p, q)}.
    \end{align*}
    \item {\em Relative distance error (RDE)} measures the per-neighbor relative distance error. Let $f(k,q,D)=(\hat p_1,\ldots,\hat p_k)$ and $\nn(k,q,D)=(p_1,\ldots,p_k)$, both ordered ascendingly w.r.t. $\dist(\cdot,q)$. Then
    \begin{align*}
        \text{RDE@}k = \frac{1}{k}\sum\nolimits_{i=1}^{k} \frac{\dist(\hat p_i,q)-\dist(p_i,q)}{\dist(p_i,q)}.
    \end{align*}
\end{itemize}

For any ANN search index and input instance, the ranges of recall, DSR, and RDE are $[0,1]$, $[1,+\infty)$, and $[0,+\infty)$, respectively. Higher recall and lower DSR/RDE values indicate better performance.

\subsection{\bb Randomly rotated \kdt{}}
\label{subsec:rrkdt-prelim}

\begin{figure}
    \centering
    \includegraphics{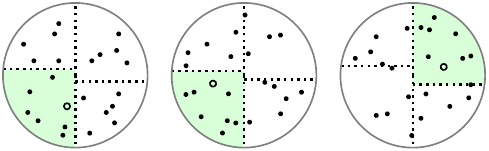}
    \caption{\bb A visualization of RR \kdt{} forest. Different rotations make the same query point $q$ fall into different leaves.}
    \label{fig: RRKDT}
\end{figure}

Our scheme is built upon the {\em randomly rotated (RR) \kdt{}}~\citep{ram2019revisiting}, a randomized variant of the classic \kdt{}~\citep{bentley1975multidimensional}. An RR \kdt{} first applies a random rotation matrix to the dataset, and applies the same rotation to each query. It then recursively partitions the rotated points along cycling coordinates, using a random split quantile from $[1/4,3/4]$ at each internal node. During query processing, it performs a {\em defeatist search}, i.e., the query follows a single root-to-leaf path without backtracking. Figure~\ref{fig: RRKDT} illustrates how different rotations send the same query to different leaf cells across trees.

We instantiate the preprocessing with the randomized circular convolution variant of Ram and Sinha~\citep{ram2019revisiting}. Specifically, $R=C_\gamma D$, where $D$ is a diagonal matrix with independent uniform signs and $C_\gamma$ is the circulant matrix generated by an independent standard Gaussian vector $\gamma$. The product $Rx$ is evaluated using FFT in $\bigo(d\log d)$ time, with the Fourier transform of $\gamma$ precomputed during construction.
% The evaluation schedule depends only on the public dimension $d$.

The defeatist search is the structural feature we rely on. Because every split leaves at most a $3/4$ fraction of the points in either child, the tree has depth $\bigo(\log N)$. Hence, each query follows a fixed-length root-to-leaf traversal, up to padding. This fixed traversal is naturally compatible with tree-based ORAM~\citep{wang2014oblivious} and is the reason we can avoid a black-box oblivious compilation of adaptive ANN probes. Although defeatist search is approximate, its accuracy is supported by its connection to {\em randomized partition trees}.

\begin{lemma}[\citep{ram2019revisiting}]
    Given any query $q\in\mathbb{R}^d$, the Recall@1 of an RR \kdt{} is identical to that of a randomized partition tree.
\end{lemma}

This theoretical equivalence places RR \kdt{}s within the broader class of ANN indexes based on randomized partitions. Prior work shows that randomized partition trees provide a competitive and often better query time and recall tradeoff than locality sensitive hashing for nearest neighbor search~\citep{sinha2014lsh}. Thus, RR \kdt{}s inherit the accuracy support of randomized partition trees while exposing the fixed path from root to leaf needed by our tree based ORAM design. Using a forest of RR \kdt{}s with independent rotations then gives multiple independent chances to retrieve true neighbors while preserving the same single path structure per tree.

\subsection{Obliviousness and ORAM}

Let $M\colon \mathcal{X}\to\mathcal{Y}$ be an algorithm expressed as a random access machine (RAM) program, $\view^M(x)$ denote the {\em access pattern} produced when $M$ runs on input $x$. This pattern records the sequence of memory locations touched by the program and may reveal information about the input even when the stored data are encrypted. Goldreich~\citep{goldreich1987towards} formalized protection against such leakage through the notion of {\em obliviousness}, formally defined as follows.

\begin{definition}[Obliviousness~\citep{goldreich1987towards}]
    Given a security parameter $\lambda$, a RAM program $M$ is {\em oblivious} if for any two inputs $x,x'\in\mathcal{X}$, the access pattern distributions $\view^M(x)$ and $\view^M(x')$ are statistically indistinguishable. That is, there exists a negligible function $\negl(\cdot)$ such that
\begin{align*}
    \Delta(\view^{M}(x), \view^{M}(x')) \leq \negl(\lambda),
\end{align*}
where $\Delta(\xi,\zeta) = \frac{1}{2}\sum_z \abs{\prob[\xi=z]-\prob[\zeta=z]}$ is the {\em statistical distance}.
\end{definition}

\paragraph{ORAM} A standard way to achieve obliviousness is to use an {\em oblivious RAM (ORAM)}. An ORAM stores an array of data blocks and supports logical read and write operations while hiding which logical addresses are accessed. In the classical client-server model, however, the ORAM may keep trusted client side states such as a position map or a stash, and accesses to this state are not required to be oblivious. The stronger notion of \emph{double obliviousness}~\citep{mishra2018oblix} removes this trusted storage assumption by requiring all memory accesses, including accesses to auxiliary state, to be oblivious.

\paragraph{Circuit ORAM} Our construction uses Circuit ORAM~\citep{wang2015circuit}, a tree based ORAM whose server side storage is a binary tree of buckets. Each logical access reads two root to leaf paths and then performs an oblivious eviction. We use the following standard bound.

\begin{lemma}[\citep{wang2015circuit}]
    \label{lem: circuitORAM}
    Let $\mathcal{C}$ be a Circuit ORAM storing blocks of size $B$ with capacity $C$ and bucket size $Z$. Then $\mathcal{C}$ supports $\mathsf{read}$ and $\mathsf{write}$ operations obliviously, each with worst case complexity $\bigo(BZ\log C)$.
\end{lemma}

A standard Circuit ORAM is not doubly oblivious because its position map and stash are client side auxiliary structures. A generic recursive construction can make these structures oblivious, but it incurs an additional logarithmic factor and yields $\bigo(\log^2 C)$ cost per logical access. Our construction avoids this black box recursion for the ORAMs used by \name{}.

\paragraph{H2O2RAM} H2O2RAM~\citep{zheng2025h2o2ram} is a hierarchical doubly oblivious RAM that organizes blocks into oblivious hash tables and periodically rebuilds its levels. Its design improves data locality and supports parallel execution of the rebuilding procedures, while protecting both internal and external memory accesses. For fixed-size blocks, it achieves an amortized access time of $\widetilde{O}(\log^2 N)$. We use H2O2RAM as an alternative storage backend for our doubly oblivious ANN baselines.

\paragraph{Oblivious sort} We also use oblivious sorting as a basic primitive. Throughout the paper, $\osort$ denotes a bitonic sorting network~\citep{batcher1968sorting}, which sorts $n$ items obliviously in $\bigo(n\log^2 n)$ (item-level) time.

\begin{figure*}[ht]
    \centering
    \includegraphics{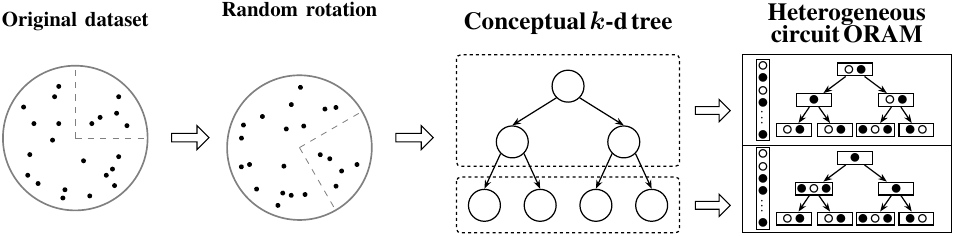}
    \caption{\bb The structure of a single oblivious \kdt{}, which first randomly rotate the original dataset, then build a conceptual \kdt{} on it, and finally map the tree to two distinct circuit ORAMs. Each ORAM contains a stash and a binary tree, where a ``\faCircleO'' represents a slot occupied by a REAL node and a ``\faCircle'' represents an empty slot (padded with a dummy node).}
    \label{fig: overview}
\end{figure*}

\section{Threat model and security goals}
\label{sec: threat model}

We consider a server equipped with a TEE that owns a database and provides a private ANN search service to multiple clients. A client's query must remain protected against a server-side adversary capable of observing all memory access patterns. These patterns reveal which memory regions are read or written and the exact sequence of those operations between the TEE-enabled processor and RAM.

To achieve strong security in our setting, \name{} provides double obliviousness~\citep{chamani2023graphos}. This property ensures that the memory access patterns remains computationally indistinguishable for any two inputs of the same size. This holds both inside the TEE's protected memory (the enclave) and outside it, including all interactions between the CPU and memory. Consequently, after processing a query, the server learns nothing beyond publicly known hyperparameters, and the client learns nothing beyond the ANN search result itself.

\paragraph{Trusted computing base (TCB)}
Our Trusted Computing Base consists solely of the TEE-enabled processor. We trust this hardware to correctly enforce confidentiality (preventing unauthorized reading of enclave contents) and integrity (preventing unauthorized modifications). We do not, however, assume that it automatically hides access patterns within enclave memory. We also rely on its remote attestation mechanism, which enables clients to verify that the enclave is genuine and executes the intended code. For instance, we assume the adversary cannot extract the hardware encryption keys used by the TEE. In other words, we do not defend against attacks that exploit hardware design flaws of the TEE implementation, such as Foreshadow~\cite{van2018foreshadow}, SGAxe~\cite{van2020sgaxe} or TEE.fail~\cite{chuang2026tee}. These attacks are orthogonal to the application-level access-pattern leakage addressed in this work. Such hardware vulnerabilities can only be mitigated through microcode patches or processor redesign.

Note that our threat model differs fundamentally from some prior TEE-based works that assume memory accesses inside the enclave are  oblivious~\citep{DBLP:journals/compsec/RachidRM20}. While both models utilize TEEs and ORAM, they typically assume an outsourced setting, where the server places the client-encrypted database outside the enclave, and achieve only single (external) obliviousness for accesses between the enclave and external storage. In contrast, our design achieves double obliviousness, delivering stronger protection overall. In fact, our techniques
can be readily adapted to support this classic outsourced setting. The client constructs the forest locally and uploads both the database and its index to the attested TEE server. Then the same doubly oblivious query execution protects query-dependent access patterns from the untrusted host.

\section{Technical overview}
\label{sec: overview}

A direct oblivious adaptation of IVF must hide cluster selection and retrieve the selected candidates through protected memory accesses. Graph-based indexes additionally involve successive expansions whose choices depend on previously evaluated candidates, limiting parallelism across search rounds. When implemented through a general-purpose doubly oblivious memory layer, these operations repeatedly incur the cost of protected address resolution and data retrieval. This motivates a search structure that simplifies both oblivious navigation and candidate access.

The cornerstone of our approach is the compatibility between defeatist search in a randomly rotated \kdt{} and pointer-based oblivious traversal. Each tree query follows one root-to-leaf path without backtracking. A parent node can therefore supply both the identifier and the ORAM position of the next node, avoiding a recursive position-map lookup~\citep{wang2014oblivious}.

Figure~\ref{fig: overview} illustrates a single tree in \name{}, where small internal nodes and large leaf buckets are placed in two ORAM trees separately. Within our scheme \name{}, a forest of $m$ such trees is stored in external memory. A query visits one leaf per tree and computes distances to all points in each visited leaf. An oblivious selection network then selects the $k$ points closest to the query from each visited leaf. Oblivious aggregation across trees produces the final answer. Candidate ranking does not alter the navigation path, so both stages have a fixed public schedule.

For a single tree, a query visits $\bigo(\log N)$ internal nodes and exactly one leaf. The internal-node accesses cost $\bigo(\log^2 N)$ in total, while fetching the leaf costs $\bigo(dn_0\log N)$. The forest repeats these accesses for all $m$ trees, with additional local selection and cross-tree aggregation costs as stated in Theorem~\ref{thm:main-informal}.

\section{\protect\boldmath \name{}: Doubly oblivious randomly rotated \kdt{} forest}
\label{sec: o2n2details}

This section provides a detailed technical description of \name{}, building upon the high-level architecture presented in Section~\ref{sec: overview}. Our two fundamental building blocks are a doubly oblivious Circuit ORAM in Section~\ref{subsec: docircuitoram} and a top-$k$ selection scheme in Section~\ref{subsec: selection-network}. We then describe single-tree queries over a doubly oblivious RR~\kdt{} in Section~\ref{subsec: tree-search}. Finally, Section~\ref{subsec: forest-query} gives the full query phase of \name{}. Table~\ref{tab:complexities} summarizes the costs of these building blocks with constant ORAM bucket size and $k\ge2$.

\begin{table}[ht]
    \centering
    \caption{Costs of ORAM and selection.}
    \label{tab:complexities}
    \begin{tabularx}{\linewidth}{lCr}
        \toprule
        \textbf{Component} & \textbf{Operation} & \textbf{Time complexity} \\
        \midrule
        Internal ORAM & Read/Write & $\bigo(\log N)$ per access \\
        Leaf ORAM & Read/Write & $\bigo(dn_0\log N)$ per access \\
        Top-$k$ selection & Select top-$k$ & $\bigo(n_0\log^2 k)$ per leaf \\
        \bottomrule
    \end{tabularx}
\end{table}

\subsection{Building block: Doubly oblivious circuit ORAM}
\label{subsec: docircuitoram}

Circuit ORAM is a natural starting point for \name{}, since a query in a randomly rotated \kdt{} already follows a root to leaf path. However, standard Circuit ORAM is not doubly oblivious, as its stash and position map are visited in a data-dependent manner. In our model, these structures reside inside the TEE enclave, so their access patterns must also be hidden.

One generic conversion is to scan the entire stash on every logical access and store the position map recursively in roughly $\bigo(\log N)$ smaller ORAMs. This supports arbitrary logical addresses, but it raises the cost of a single logical ORAM access from $\bigo(\log N)$ to $\bigo(\log^2 N)$. Since \name{} accesses tree nodes through this ORAM interface, the extra logarithmic factor would propagate to the costs.

We instead use position embedding~\citep{wang2014oblivious}, i.e., let each parent node in the conceptual \kdt{} store the identifiers and ORAM positions of its children. Reaching a child through its parent provides the required position directly. Stash operations scan a fixed-size array, using masked moves to select or update entries. The eviction follows the oblivious Circuit ORAM schedule~\citep{wang2015circuit}. With a stash of $\bigo(\log N)$ blocks configured for the desired failure probability, the pointer-based interface costs $\bigo(BZ\log N)$ per access, where $B$ is the data block size and $Z$ is the ORAM bucket size.

Let $\mathcal{T}$ be a Circuit ORAM of $\bigo(N)$ capacity. We address each block by a pointer $\ptr=(\idx,\pos)$ containing its identifier and ORAM position. The ORAM provides the following interfaces.
\begin{itemize}[leftmargin=*]
    \item $\mathcal{T}.\mathsf{ReadRm}(\ptr)$ scans the stash and the ORAM path identified by $\ptr.\pos$, returns the block with key $\ptr.\idx$, and replaces it by a dummy block $\dummy$.
    \item $\mathcal{T}.\mathsf{Add}(\ptr,x)$ scans the stash and places block $x$ with pointer $\ptr$ in a dummy slot using masked updates.
    \item $\mathcal{T}.\mathsf{Evict}()$ runs the public, data-independent Circuit ORAM eviction procedure~\citep{wang2015circuit}.
\end{itemize}

We write $\mathsf{Write}$ for $\mathsf{Add}$ followed by $\mathsf{Evict}$. The query schedule explicitly alternates removal and reinsertion, so a block can be updated between its $\mathsf{ReadRm}$ and $\mathsf{Write}$ calls without an intermediate dummy reinsertion. In \name{}, the read/write schedule is fixed independently of the query, so distinguishing reads from writes reveals no query-dependent information.

\subsection{Building block: Selection network}
\label{subsec: selection-network}

Our second building block selects the $k$ closest candidates to a query $q$ from an array of $n_0$ candidates stored in a single \kdt{} leaf. Here, each candidate is represented by a constant-size record $(\delta,\idx)$ containing its distance to $q$ and its identifier. We order records first by distance and then by identifier, with dummy records placed after all valid records. This order preserves Euclidean distance ranking and resolves ties consistently throughout the query.

We use the combined selection network of Cong et al.~\citep{cong2025revisiting}. It combines truncated odd-even merging with Yao's recursive selection. The former merges sorted candidate lists while omitting comparisons unnecessary for the top-$k$. The latter compares disjoint pairs and recursively retains the best $\lfloor k/2\rfloor$ records from the larger members. It then selects the top-$k$ from these records, the smaller members, and any unpaired record. The combined construction chooses between these approaches according to their comparator counts for the public subproblem sizes. It returns the $k$ best records without requiring them to be sorted.

We generate the comparator schedule before querying. All comparisons and exchanges use fixed record addresses and masked moves, so the access schedule depends only on leaf size $n_0$ and output size $k$. The combined network uses no more comparisons than truncated selection alone, giving $\bigo(n_0\log^2 k)$ time for $k\ge2$. For $k=1$, a minimum tournament takes linear time.

\begin{remark}
    A max heap of capacity $k$ is a standard way to retain the best candidates, with the worst retained candidate at the root. A doubly oblivious max heap is therefore an alternative for this building block. We construct a restricted DO heap based on circuit ORAM, incurring $\bigo(\log^2 c)$ cost for Insert and ExtractMin operations. Within our scope, it is slightly faster than Shi's path oblivious heap~\citep{shi2020path}. We describe its construction in Appendix~\ref{app:heap} and compare it with a selection network in the ablation in Figure~\ref{fig:selection-ablation}.
\end{remark}

\subsection{\protect\boldmath Doubly oblivious randomly rotated \kdt{}}
\label{subsec: tree-search}

We represent a single conceptual randomly rotated (RR) \kdt{} by a tuple $(\mathcal{T}_{\it int},\mathcal{T}_{\it leaf},\root,R,h)$. Here $R$ is the tree's fast structured transform from Section~\ref{subsec:rrkdt-prelim}, $\root$ is its root pointer, and $h=\bigo(\log N)$ is the public number of internal levels, with shorter paths padded to this depth. The internal-node ORAM $\mathcal{T}_{\it int}$ stores blocks $b=(a,\tau,\lptr,\rptr)$, where $a\in\{1,\ldots,d\}$ is the splitting coordinate in the transformed space, $\tau$ is the split threshold, and $\lptr$ and $\rptr$ are the left and right child pointers, with form $(\idx,\pos)$ defined in Section~\ref{subsec: docircuitoram}. Each block in the leaf ORAM $\mathcal{T}_{\it leaf}$ stores a bundle of points from a single \kdt{} leaf, padded with dummies to $n_0$ slots. Points retain their original coordinate representation for distance evaluation.

Separating the two node types avoids padding every tiny $\mathcal{T}_{\it int}$ block to $\Theta(dn_0)$ size. A query visits $h$ internal nodes and one leaf. A unified ORAM would charge $\bigo(dn_0Z\log N)$ per internal access. Separate ORAMs reduce the ORAM access cost to $\bigo(Z\log N)$ per internal node and pay the large-block cost only for the leaf.

For a query $q\in\R^d$, Algorithm~\ref{alg: orrkdt-query} computes $q'=R\cdot q$ using FFT and follows one root-to-leaf path in the conceptual \kdt{}. At each internal node, $\alpha=(q'_a\leq\tau)$ selects $\lptr$ when true and $\rptr$ otherwise, yielding $\ptr_{\mathsf{next}}$. The access to $q'_a$ is implemented by scanning all $d$ coordinates with masked moves to keep oblivious. Pointer selection and replacement likewise access both child fields on a fixed schedule, changing only the selected field during replacement.

At the start of each iteration, $\ptr$ identifies the current block at its existing ORAM position, while $\pos_{\mathsf{out}}$ is its fresh reinsertion position, already recorded in $\root$ or its parent. After removing $b$ with $\mathsf{ReadRm}(\ptr)$, the algorithm samples $\pos_{\mathsf{next}}$ for the selected child and records $(\ptr_{\mathsf{next}}.\idx,\pos_{\mathsf{next}})$ in $b$. It then reinserts $b$ at $(\ptr.\idx,\pos_{\mathsf{out}})$ using $\mathsf{Write}$. Retaining the child's existing pointer in $\ptr_{\mathsf{next}}$ allows the next iteration to read that child before reinserting it at $\pos_{\mathsf{next}}$. Each fresh position is sampled uniformly from the public position range of the ORAM storing the corresponding node. After exactly $h$ internal iterations, the algorithm reads the leaf block in $\mathcal{T}_{\it leaf}$ containing $S_q$ and reinserts it unchanged at its reserved position. In other words, every query performs $h$ internal-node read/write pairs and one leaf read/write pair. Finally, it calls the selector from Section~\ref{subsec: selection-network} (denoted by $\mathsf{Select}_k(S_q,q)$) to retain the leaf's top-$k$ closest candidates and returns them as $S_q$. 

\begin{algorithm}[ht]
    \caption{Query of oblivious \kdt{}}
    \label{alg: orrkdt-query}
    \KwIn{Index $(\mathcal{T}_{\it int},\mathcal{T}_{\it leaf},\root,R,h)$, Query $q\in\R^d$}
    \KwOut{Top-$k$ candidates $S_q$ from the visited leaf, padded with dummies if necessary}
    $q'\gets R\cdot q$\;
    $\ptr\gets\root$\;
    Sample $\pos_{\mathsf{out}}$ uniformly at random and update $\root\gets(\root.\idx,\pos_{\mathsf{out}})$\;
    \For{$\ell\gets 1$ \KwTo $h$}{
        $b=(a,\tau,\lptr,\rptr)\gets \mathcal{T}_{\it int}.\mathsf{ReadRm}(\ptr)$\;
        $\alpha\gets(q'_a\leq \tau)$\;
        $\ptr_{\mathsf{next}}\gets$ the child pointer selected by $\alpha$\;
        Sample $\pos_{\mathsf{next}}$ uniformly at random\;
        In $b$, replace the selected child pointer by $(\ptr_{\mathsf{next}}.\idx,\pos_{\mathsf{next}})$\;
        $\mathcal{T}_{\it int}.\mathsf{Write}((\ptr.\idx,\pos_{\mathsf{out}}),b)$\;
        $\ptr\gets\ptr_{\mathsf{next}}$\;
        $\pos_{\mathsf{out}}\gets\pos_{\mathsf{next}}$\;
    }
    $S_q\gets \mathcal{T}_{\it leaf}.\mathsf{ReadRm}(\ptr)$\;
    $\mathcal{T}_{\it leaf}.\mathsf{Write}((\ptr.\idx,\pos_{\mathsf{out}}),S_q)$\;
    $S_q\gets\mathsf{Select}_k(S_q,q)$\;
    \KwRet{$S_q$}
\end{algorithm}

\subsection{Query phase of \name{}}
\label{subsec: forest-query}

We now describe the full query phase of \name{}, which composes the single-tree query procedure with oblivious candidate aggregation. The input is a forest of $m$ doubly oblivious RR \kdt{}s with independent transforms. For a query $q$, the algorithm queries every tree using Algorithm~\ref{alg: orrkdt-query}. Each tree returns $k$ candidate slots selected from its visited leaf.

The candidates returned by all trees are appended to an array $\mathcal{A}$. After oblivious duplicate removal, a final selection network produces the best $k$ distinct candidates as $S_q$. The duplicate-removal procedure sorts $\mathcal{A}$ by identifier and replaces repeated identifiers by dummies in a fixed scan, preserving the array length.

\begin{algorithm}[ht]
    \caption{Approximate $k$-NN query of \name{}}
    \label{alg: forest-query}
    \KwIn{Index $\mathcal{F}=(\mathcal{F}_1,\mathcal{F}_2,\ldots,\mathcal{F}_m)$, Query $q\in\R^d$}
    \KwOut{Approximate $k$-NN set $S_q$}

    $\mathcal{A}\gets\varnothing$\;
    \For{$i\gets 1$ \KwTo $m$}{
        $(s_1,s_2,\ldots,s_k)\gets\mathcal{F}_i.\mathsf{query}(q,k)$\;
        \For{$j\gets 1$ \KwTo $k$}{
            Append $(s_j,\dist(s_j,q))$ to $\mathcal{A}$\;
        }
    }
    $\mathcal{A}\gets \osort(\mathcal{A})$\;
    Replace all duplicate items in $\mathcal{A}$ with $\dummy$\;
    $S_q\gets\mathsf{Select}_k(\mathcal{A},q)$\;
    \KwRet{$S_q$}
\end{algorithm}

\section{Correctness and obliviousness}
\label{sec: guarantee}

This section proves the guarantees behind Theorem~\ref{thm:main-informal}. We first establish correctness and double obliviousness for the single-tree query, then compose these facts to obtain the guarantee for \name{}. The running-time bound follows from the public loop bounds in Algorithm~\ref{alg: forest-query} and the costs in Table~\ref{tab:complexities}.

\paragraph{Correctness} We show that Algorithm~\ref{alg: orrkdt-query} simulates plaintext defeatist search followed by top-$k$ selection in the visited leaf.

\begin{claim}[Correctness of single-tree query]
    \label{claim: okdt-correct}
Let $T$ be a plaintext RR \kdt{} with transform $R$ and padded depth $h$. Suppose that $(\mathcal{T}_{\it int},\mathcal{T}_{\it leaf},\root,R,h)$ stores the same logical tree as $T$ in the following sense.
\begin{enumerate}[leftmargin=*]
    \item The root pointer $\root$ identifies the root node of $T$ at its current ORAM position.
    \item Each internal block stores the same split coordinate, threshold, and logical child identifiers as the corresponding node of $T$, with valid ORAM positions in both child pointers.
    \item Each leaf block stores the points in the corresponding leaf of $T$, padded with dummies to $n_0$ slots.
\end{enumerate}
Then, for every query $q$, Algorithm~\ref{alg: orrkdt-query} returns the top-$k$ points from the leaf reached by plaintext defeatist search on $(T,R)$, using the distance--ID order from Section~\ref{subsec: selection-network} and dummy padding when necessary. The logical tree and pointer consistency are preserved.
\end{claim}

\begin{proof}
Let $q'=R\cdot q$. We use a loop invariant over the $h$ internal-node rounds. At the beginning of each round, $\ptr$ locates the same logical node $u$ as the plaintext query, while $\pos_{\mathsf{out}}$ is the fresh position reserved for $u$ in its root or parent pointer. All other stored pointers remain valid, and the logical tree is unchanged. The invariant holds initially after the algorithm reserves the root's new position.

Consider one round at $u$. The block read from $\mathcal{T}_{\it int}$ contains the same splitting coordinate $a$, threshold $\tau$, and child identifiers as $u$ in $T$. The coordinate scan returns $q'_a$, so $q'_a\leq\tau$ selects the same child as the plaintext query. The algorithm retains that child's current pointer in $\ptr_{\mathsf{next}}$ and records its fresh position $\pos_{\mathsf{next}}$ in the parent block. Writing the parent at $\pos_{\mathsf{out}}$ restores its incoming pointer's validity. Advancing to $\ptr_{\mathsf{next}}$ with reserved position $\pos_{\mathsf{next}}$ therefore preserves the invariant.

After $h$ rounds, $\ptr$ identifies the same leaf as the plaintext query. Its $\mathsf{ReadRm}$ returns the same point set, and its $\mathsf{Write}$ restores the last pending pointer without changing the leaf contents. By the correctness of the selection network of Cong et al.~\citep{cong2025revisiting}, the final $\mathsf{Select}_k$ returns that leaf's top-$k$ according to distances to the original query $q$, padded with dummies if necessary.
\end{proof}

\paragraph{Double obliviousness} A single-tree query hides both the selected traversal path and the subsequent candidate selection.

\begin{claim}[Double obliviousness of single-tree query]
    \label{claim: okdt-oblivious}
Let $(\mathcal{T}_{\it int},\mathcal{T}_{\it leaf},\root,R,h)$ be an RR \kdt{} stored using the pointer-based ORAM interface in Section~\ref{subsec: docircuitoram}. For fixed public parameters, the query procedure in Algorithm~\ref{alg: orrkdt-query} is doubly oblivious under the security guarantee of that interface.
\end{claim}

\begin{proof}
Fix the public parameters and consider any two valid query--index states sharing these parameters. In both executions, the algorithm performs exactly $h$ internal rounds on $\mathcal{T}_{\it int}$, each with one $\mathsf{ReadRm}$ and one $\mathsf{Write}$, followed by one leaf round on $\mathcal{T}_{\it leaf}$ with the same operation types. The resulting logical access sequences have length $t=2h+2$ and agree in their ORAM instances and operation types at every position. Their logical pointers may differ adaptively. Fresh positions are sampled independently, and the doubly oblivious ORAM interface hides these logical choices, including accesses to the stash. Its sequential security guarantee therefore gives indistinguishable physical traces for the two access sequences.

Outside the ORAM calls, the transform, coordinate scans, and masked pointer updates follow fixed schedules. Distance evaluation scans all $n_0$ point slots, and the selection network uses public comparator addresses with masked exchanges and returns exactly $k$ slots. These local traces depend only on public parameters. Composing them with the ORAM accesses preserves indistinguishability of the complete traces, proving double obliviousness.
\end{proof}

\paragraph{Main guarantee} We now combine the two claims to obtain the formal version of Theorem~\ref{thm:main-informal}.

\begin{theorem}[Guarantee of \name{}]
    \label{thm:o2n2-guarantee}
For $N$ points in $\R^d$ with $d\leq N$, $m$ trees and leaf capacity $n_0$, Algorithm~\ref{alg: forest-query} correctly simulates the plaintext RR \kdt{} forest query with the same trees, transforms, and distance--ID order, returning the best $k$ distinct points in the visited leaves and satisfying double obliviousness. Using constant ORAM bucket size, its worst-case query time is
\begin{align*}
    \bigo\bigl(m\log^2 N+mn_0d\log N+mk\log^2(mk)\bigr).
\end{align*}
\end{theorem}

\begin{proof}
For correctness, Claim~\ref{claim: okdt-correct} shows that each single-tree query returns the top-$k$ of the same leaf as plaintext defeatist search. Retaining only these points is lossless for global top-$k$ selection. Every discarded point has at least $k$ distinct points preceding it in the same leaf under the common distance--ID order, so it cannot belong to the top-$k$ of the union of visited leaves. Sorting the retained candidates by identifier makes duplicates adjacent, and the fixed scan replaces repeated occurrences by dummies. The final $\mathsf{Select}_k$ therefore returns the best $k$ distinct candidates, exactly matching the plaintext forest query.

For double obliviousness, Algorithm~\ref{alg: forest-query} makes a public number $m$ of single-tree queries, each doubly oblivious by Claim~\ref{claim: okdt-oblivious}, and appends exactly $k$ candidate records per tree. Aggregation uses oblivious sorting, a fixed scan for duplicate marking, and a selection network on the resulting $mk$ slots. Their access schedules depend only on public parameters, so their composition with the tree queries is doubly oblivious.

For running time, Table~\ref{tab:complexities} gives $\bigo(\log^2 N)$ for the $\bigo(\log N)$ internal-node accesses per tree and $\bigo(dn_0\log N)$ for the leaf access. The local selector costs $\bigo(n_0\log^2 k)$ for $k\ge2$. Since $d\leq N$, the transform, coordinate scans, and distance evaluation are bounded by the leaf-access cost, as is the linear selector for $k=1$. Multiplying the per-tree costs by $m$ gives the first two terms. Aggregation over the $mk$ retained candidates takes $\bigo(mk\log^2(mk))$ time for $mk\ge2$, and constant time otherwise, giving the last term.
\end{proof}

\section{Experiments}

\begin{figure*}[htbp]
    \centering
    \includegraphics{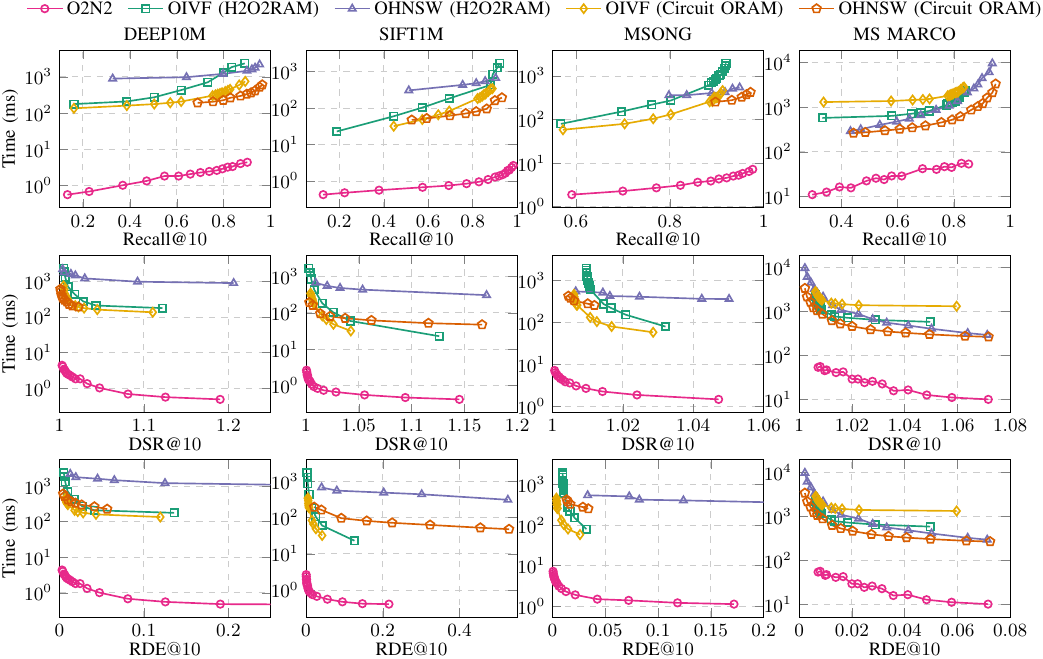}
    \caption{\bf\boldmath Query time and quality for $10$-NN search.}
    \label{fig: tradeoff-k10}
\end{figure*}

\begin{figure*}[htbp]
    \centering
    \includegraphics{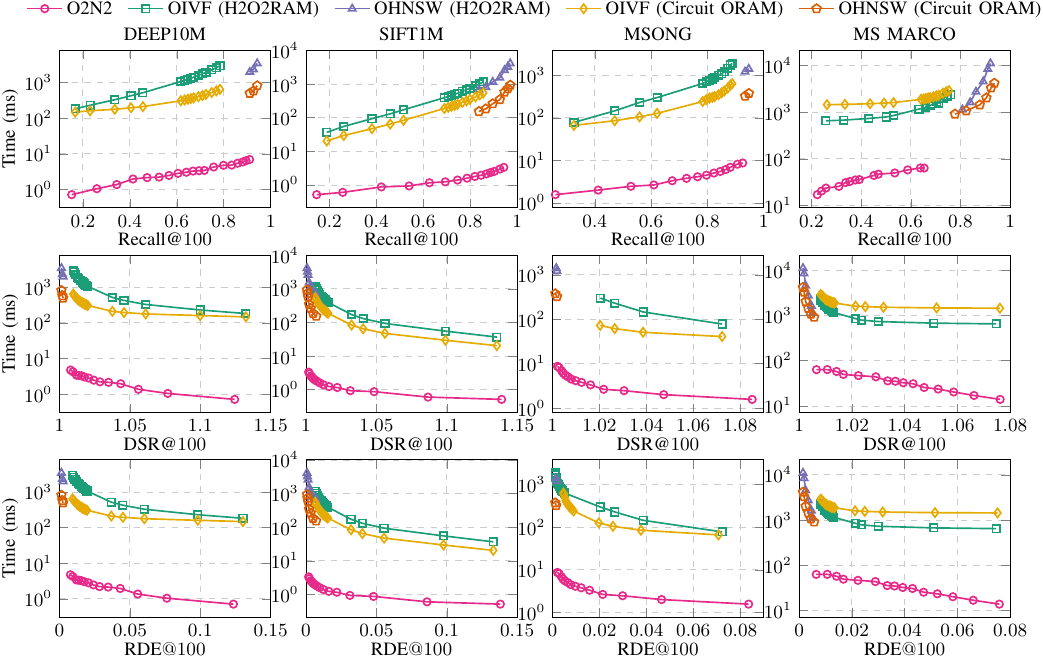}
    \caption{\bf\boldmath Query time and quality for $100$-NN search.}
    \label{fig: tradeoff-k100}
\end{figure*}

In this section, we evaluate the performance of our proposed scheme, \name{}, and compare it against several doubly oblivious baselines. We implement all algorithms in C++20 and evaluate them on the benchmark datasets summarized in Table~\ref{tab: workload-stats}. We additionally report O2N2 scalability results on SIFT100M, the first 100 million base vectors of BIGANN, using its public queries and 100M ground truth~\citep{simhadri2022results}.

\paragraph{Workload difficulty} To summarize workload difficulty, we use two metrics, local intrinsic dimensionality (LID) and relative contrast (RC). Given dataset $D$ and query $q$, LID and RC are defined as
\begin{align*}
    \mathrm{LID}(q,D)=\lim_{r\to 0}\frac{rF_q'(r)}{F_q(r)}, ~~
    \mathrm{RC}(q,D)=\frac{\mathbb{E}_{p}[\dist(q,p)]}{r_1(q)},
\end{align*}
where $F_q(r)$ denotes the probability that a database point lies within distance $r$ of query $q$, and $r_i(q)$ denotes the distance between $q$ and the $i$-th NN. Following Aum{\"u}ller and Ceccarello~\citep{aumuller2019role}, we use the same $100$-NN maximum-likelihood estimator for LID, and estimate RC by a sample average on $S$ with size $10^5$.
\begin{align*}
    \widehat{\mathrm{LID}} = \left(\frac{1}{99}\sum\limits_{i=1}^{99}\log\frac{r_{100}(q)}{r_i(q)}\right)^{-1}, \widehat{\mathrm{RC}} = \frac{\sum_{p\in S}\dist(q,p)}{|S|\cdot r_1(q)}.
\end{align*}
Table~\ref{tab: workload-stats} reports averages over all queries. Higher LID values and lower RC values generally indicate more challenging workloads. Among these datasets, DEEP10M has the highest LID, while MS MARCO has the lowest RC.

\begin{table}[htbp]
    \centering
    \caption{Datasets and workload statistics.}
    \label{tab: workload-stats}
    \setlength{\tabcolsep}{2pt}
    \begin{tabular*}{\linewidth}{@{\extracolsep{\fill}}lrrrrrr@{}}
        \toprule
        \multirow{2}{*}{\bf Dataset} & \multicolumn{1}{c}{\multirow{2}{*}{\bb $N$}} & \multicolumn{1}{c}{\multirow{2}{*}{\bb $d$}} & \multicolumn{2}{c}{\bf LID} & \multicolumn{2}{c}{\bf RC} \\
        % \cmidrule(lr){4-5}\cmidrule(l){6-7}
        & & & mean & median & mean & median \\
        \midrule
        DEEP10M~\citep{aumuller2020ann} & 9,990,000 & 96 & {\bb 25.22} & {\bb 25.20} & 3.41 & 2.40 \\
        SIFT1M~\citep{aumuller2020ann} & 1,000,000 & 128 & 21.70 & 21.67 & 3.19 & 2.75 \\
        MSONG~\citep{bertin2011million} & 994,185 & 420 & 23.83 & 23.46 & 3.98 & 3.48 \\
        MS MARCO~\citep{bajaj2016msmarco,lin2021pyserini} & 8,841,823 & 768 & 6.40 & 6.23 & {\bb 1.84} & {\bb 1.81} \\
        \bottomrule
    \end{tabular*}
\end{table}
Our experiments are conducted on a confidential machine protected by Intel\textregistered{} Trust Domain Extensions (TDX)~\citep{intel-tdx-whitepaper-2022}. The underlying physical host is powered by an Intel Xeon Platinum 8469C processor (codenamed Sapphire Rapids). The guest VM is configured with 192 vCPUs operating at 2.6 GHz, and allocated $3\,\text{TiB}$ of RAM. We use 32 query worker threads for \name{}, with different trees searched in parallel. All experiments are performed over a local network\footnote{Since our protocol requires only a single round trip with tens of kilobytes of communication between the client and server, readers can directly add their wide-area network RTT to estimate performance in geographically distributed settings.} with 5Gbps bandwidth and sub-millisecond round-trip latency.

\subsection{Baselines}
\label{subsec: baselines}

Since no prior work addresses doubly oblivious ANN, our only viable baselines are strengthing popular plaintext ANN algorithms  with existing doubly oblivious RAMs. We construct baselines from two standard ANN indexes, Inverted File Index (IVF) and Hierarchical Navigable Small World (HNSW).

\paragraph{Doubly oblivious IVF index (OIVF)} A typical IVF index consists of a cluster collection $\boldsymbol{D}$ and point storage. Here $\boldsymbol{D}=(D_1,\ldots,D_m)$ contains $m$ clusters, and each cluster $D_i$ contains its centroid and a point list assigned to that centroid. OIVF stores point records in a doubly ORAM $\mathcal{O}$ and pads every point list to the public capacity $L_{\max}$. Given a query point $q$, it sorts $\boldsymbol{D}$ by centroid distance to $q$, fetches $L_{\max}$ points from each of the first $t$ clusters, and returns the best $k$ candidates after an oblivious sort. The query details are shown in Algorithm~\ref{alg: oivf-query}, of which the complexity is $\bigo(dm+m\log^2m+tL_{\max}(d\log^2N+\log^2(tL_{\max})))$, where we assume each access of $\mathcal{O}$ incurs $\bigo(\log^2 N)$ runtime. In our experiments, we set target cluster size to $25$, which is a public constant. Since a cluster can exceed its target after clustering, we pad each point list to $L_{\max}=50$.

\begin{algorithm}[t]
    \caption{Approximate $k$-NN query of OIVF}
    \label{alg: oivf-query}
    \KwIn{Index $(\mathcal{O},m,\boldsymbol{D},t,L_{\max})$, Query $q\in \R^d$}
    \KwOut{Approximate $k$-NN set $S_q$}

    $S_q\gets \varnothing$\;
    $\osort(\boldsymbol{\boldsymbol{D}})$ ascendingly by distance between centroids and $q$\;
    \For{$i\gets 1$ \KwTo $t$}{
        $S\gets$ fetch the $L_{\max}$ points in $D_i$ from $\mathcal{O}$\;
        $S_q\gets S_q\cup S$\;
    }
    $\osort(S_q)$ by the distance to $q$ in ascending order\;
    $S_q\gets$ first $k$ elements of $S_q$\;
    \KwRet{$S_q$}
\end{algorithm}

% \begin{algorithm}[t]
%     \caption{Query phase of OIVF}
%     \label{alg: oivf-query}
%     \KwIn{OIVF index $(\mathcal{O}, C, \boldsymbol{D}, t, L_{\max})$, query $q\in \R^d$, size $k$}
%     \KwOut{Approximate $k$-NN set $S_q$}

%     $A\gets \{(\dist(q,c_i),D_i): c_i\in C\}_{i=1}^{m}$\;
%     $A\gets \osort(A)$ in ascending distance order\;
%     $P\gets \varnothing$\;
%     \For{$r\gets 1$ \KwTo $t$}{
%         $(\cdot,D_r)\gets A[r]$\;
%         \For{$j\gets 1$ \KwTo $L_{\max}$}{
%             $x\gets \mathcal{O}.\mathsf{Read}(D_r[j])$\;
%             $P\gets P\cup \{(\dist(q,x),D_r[j])\}$\;
%         }
%     }
%     $P\gets \osort(P)$ in ascending distance order\;
%     $S_q\gets$ ids of the first $k$ non-dummy entries in $P$\;
%     \KwRet{$S_q$}
% \end{algorithm}

\paragraph{Doubly oblivious HNSW index (OHNSW)} A typical HNSW index consists of an entry point $s$, a layered graph $\mathcal{G}$, and point storage $\mathcal{O}$. OHNSW stores point records in $\mathcal{O}$ and graph adjacency arrays in $\mathcal{G}$. The graph has top layer $\ell$ and public degree bounds $\{\Delta_l\}_{l=0}^{\ell}$. Given a query point $q$, query processing starts from $s$, uses $\rho$ fixed greedy rounds on each upper layer, and then runs a base layer beam search with public beam width $w$ and fixed visit budget $\tau$. The query details are shown in Algorithm~\ref{alg: ohnsw-query}, which returns the final $k$ approximate nearest neighbors. Its ORAM-dominated query complexity is $\bigo((\rho\ell +\tau)\cdot d\Delta\log^2 N)$, where we assume each backed ORAM access incurs $\bigo(\log^2 N)$ runtime and let $\Delta=\max\{\Delta_0,\Delta_1,\ldots, \Delta_\ell\}$. In our experiments, DEEP10M and MS MARCO use upper-layer degree bounds $\Delta_1=\cdots=\Delta_\ell=32$ and base-layer degree bound $\Delta_0=64$, while the other datasets use upper-layer degree bounds $\Delta_1=\cdots=\Delta_\ell=16$ and base-layer degree bound $\Delta_0=32$.

\begin{algorithm}[t]
    \caption{Approximate $k$-NN query of OHNSW}
    \label{alg: ohnsw-query}
    \KwIn{Index $(\mathcal{O}, \mathcal{G}, s, \ell, \{\Delta_l\}_{l=0}^{\ell}, \rho, w, \tau)$, query $q\in \R^d$}
    \KwOut{Approximate $k$-NN set $S_q$}

    $e\gets s$\;
    \For{$l\gets \ell$ \KwTo $1$}{
        \For{$r\gets 1$ \KwTo $\rho$}{
            $v\gets \mathcal{O}.\mathsf{Read}(e)$\;
            $\mathcal{N}\gets$ padded neighbor ids from $\mathcal{G}.\mathsf{Read}(e,l)$\;
            $e\gets$ id in $\{e\}\cup \mathcal{N}$ closest to $q$\;
        }
    }

    $v\gets \mathcal{O}.\mathsf{Read}(e)$\;
    Initialize arrays $R$, $C$, and $V$:
    \begin{enumerate}[leftmargin=*]
        \item Fill $R$ and $C$ with $(\dist(q,v),e)$ and $w-1$ copies of $(+\infty,\dummy)$\;
        \item Fill $V$ with $e$ and $\tau\cdot\Delta_0$ copies of $\dummy$\;
    \end{enumerate}
    
    \For{$r\gets 1$ \KwTo $\tau$}{
        $(d,j)\gets \arg\min_{(dis,id)\in C} dis$\;
        $C\gets C\setminus\{(d,j)\}\cup\{(+\infty,\dummy)\}$\;
        $\mathcal{N}\gets$ padded neighbor ids from $\mathcal{G}.\mathsf{Read}(j,0)$\;
        \For{$id\in \mathcal{N}$}{
            $u\gets \mathcal{O}.\mathsf{Read}(id)$\;
            \If{$id\neq\dummy$ and $id\notin V$}{
                Add $id$ to $V$ by a linear scan\;
                $(d_{\max},id_{\max})\gets \arg\max_{(dis,id')\in C} dis$\;
                \If{$\dist(q,u)<d_{\max}$}{
                    $C\gets C\setminus\{(d_{\max},id_{\max})\}\cup\{(\dist(q,u),id)\}$\;
                }
                Update $R$ with $(\dist(q,u),id)$ by a linear scan\;
            }
        }
    }
    $R\gets \osort(R)$ in ascending distance order\;
    $S_q\gets$ ids of the first $k$ non-dummy entries in $R$\;
    \KwRet{$S_q$}
\end{algorithm}

\paragraph{Double obliviousness of the baselines}

In Algorithms~\ref{alg: oivf-query} and~\ref{alg: ohnsw-query}, we implement minimum/maximum selection and membership tests using full scans, conditionals and set operations using masked updates on fixed-capacity arrays, and sorting using oblivious networks. Loop bounds and array lengths are public, and each query returns exactly $k$ slots. Dummy accesses use reserved ORAM records with padded empty adjacency lists where needed. Intuitively, the doubly ORAMs hide the logical addresses accessed during search, while the fixed query schedule and oblivious local operations hide data-dependent choices. Formally, we have the following proposition.

\begin{proposition}[Double obliviousness of Baselines]
\label{prop:baseline-oblivious}
Given OIVF and OHNSW indexes, Algorithms~\ref{alg: oivf-query} and~\ref{alg: ohnsw-query} are doubly oblivious.
\end{proposition}
\begin{proof}
Throughout the proof, we assume that all public parameters are fixed.

For OIVF, centroid distance evaluation and oblivious sorting process all $m$ fixed-width cluster records, each containing a centroid and its padded identifier list. The algorithm then scans the first $t$ records at fixed offsets and reads all $L_{\max}$ point slots per cluster, including dummies. It therefore issues exactly $tL_{\max}$ ORAM reads, whose logical addresses depend on the selected clusters. Distance evaluation, final oblivious sorting, and output of the first $k$ slots follow fixed local access schedules.

For OHNSW, the algorithm executes $\rho$ rounds per upper layer and $\tau$ base-layer rounds. Each round processes the full padded neighbor list. These fixed rounds and list lengths, together with the scheduled entry-point reads, determine the number and order of graph and point ORAM accesses. Candidate selection, membership tests, and updates scan the fixed-capacity arrays $R,C,V$ using masked operations, even for dummy, visited, or rejected candidates. Final oblivious sorting and output of the first $k$ slots also follow fixed schedules.

Thus, each algorithm is a fixed-schedule composition of local oblivious computations and doubly ORAM calls. Query-dependent choices affect the logical addresses supplied to ORAM, but not the call counts, operation types, or local access schedules. By the security of the underlying doubly ORAMs against adaptive logical accesses, these calls have indistinguishable internal and external access traces for any two valid inputs sharing the public parameters. Their composition with the local oblivious computations therefore proves double obliviousness of both algorithms.
\end{proof}

\paragraph{Choosing ORAM backend}

We instantiate two baselines using different doubly oblivious RAM backends. The first is built on our own implementation of recursive Circuit ORAM~\citep{wang2015circuit}. This serves as a natural reference point in the doubly oblivious setting: when client-side computation must itself be oblivious, its eviction circuit avoids the additional $\log\log N$ factor incurred by other state-of-the-art ORAMs such as Path ORAM~\citep{stefanov2018path}. The second baseline uses the open-source implementation of H2O2RAM~\citep{zheng2025h2o2ram}, which was reported to significantly outperform EnigMap~\citep{tinoco2023enigmap} and GraphOS~\citep{chamani2023graphos}.

\subsection{Experimental results}
\begin{figure}[htbp]
    \centering
    \includegraphics{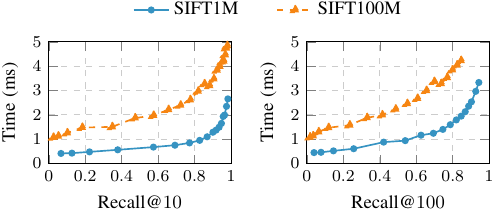}
    \caption{\boldmath Query time versus recall of \name{} on SIFT1M and SIFT100M for $k=10$ and $k=100$.}
    \label{fig: sift100m}
\end{figure}

The main baseline comparisons cover four datasets: DEEP10M, SIFT1M, MSONG, and MS MARCO (Table~\ref{tab: workload-stats}). We additionally evaluate \name{} on SIFT100M in Figure~\ref{fig: sift100m} and Table~\ref{tab: results on high recalls}. For \name{}, we set the leaf size to $n_0 = \lceil\log^2 N\rceil$ and the ORAM bucket size to $Z = 4$ across all datasets. The number of trees $m$ is varied to trace the time-recall tradeoff curves. Our primary metrics are query time, recall accuracy ($\text{Recall}@k$), distance sum ratio ($\text{DSR}@k$), and RDE ($\text{RDE}@k$) for $k=10$ and $k=100$.

\paragraph{Overall performance tradeoffs} 
Figures~\ref{fig: tradeoff-k10} and~\ref{fig: tradeoff-k100} report the query time\footnote{Interestingly,  Circuit ORAM baselines outperform H2O2RAM baselines in many settings, we observe that this is because  H2O2RAM's efficiency depends heavily on cache locality  across long sequences of consecutive accesses, while the memory-intensive ANN computations interleaved with ORAM accesses inevitably break this locality and pollute the cache.} versus recall, DSR, and RDE for the $10$-NN and $100$-NN results. In the reported regimes, \name{} generally gives the best latency.
The performance gap comes from how each index maps to oblivious execution. To reach high recall, OIVF scans many candidate points from the selected clusters, while OHNSW performs a fixed-budget graph traversal that still follows a sequential dependency chain. Each  step depends on the previous step, creating serialized ORAM accesses that are difficult to parallelize. 
In contrast, \name{} queries $m$ independent \kdt{}s. These tree queries can run concurrently,  allowing \name{} to turn additional parallel resources into better accuracy.  

\paragraph{Scalability on SIFT100M}
Figure~\ref{fig: sift100m} compares the time--recall tradeoffs on SIFT1M and SIFT100M as the dataset size increases by a factor of $100$. With the forest size fixed at $m=64$, measured query latency increases by factors of $2.17$ and $1.67$ for $k=10$ and $k=100$, respectively. The same forest size also yields lower recall on SIFT100M for both values of $k$. Scaling to the larger dataset therefore increases the cost at a fixed tree count and requires more trees to attain comparable recall.

\begin{figure}[htbp]
    \centering
    \includegraphics{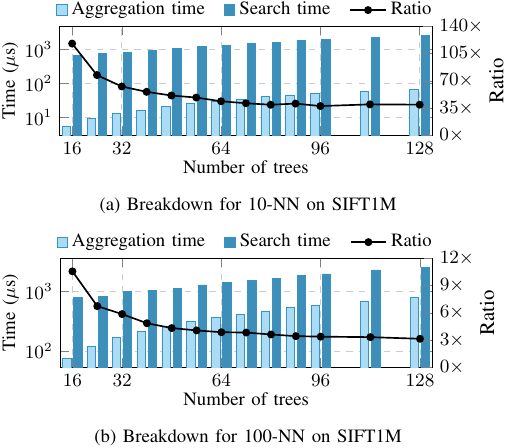}
    \caption{\boldmath Query-time breakdown of \name{} on SIFT1M for 10-NN and 100-NN. For both panels, the right axes show the search-to-aggregation ratio.}
    \label{fig: breakdown}
\end{figure}

\begin{figure}[htbp]
    \centering
    \includegraphics{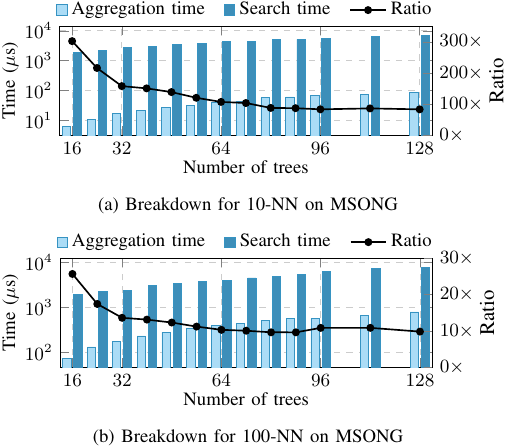}
    \caption{\boldmath Query-time breakdown of \name{} on MSONG for 10-NN and 100-NN. For both panels, the right axes show the search-to-aggregation ratio.}
    \label{fig: breakdown-msong}
\end{figure}

\paragraph{Breakdown of query time}
Figures~\ref{fig: breakdown} and~\ref{fig: breakdown-msong} separate query latency on SIFT1M and MSONG into forest search and cross-tree aggregation for $k=10$ and $k=100$. The search time includes the query transform, ORAM navigation and leaf access, distance computation, and local selection, with the trees queried in parallel. Aggregation time consolidates the returned candidates. The overlaid line reports the search-to-aggregation ratio. On SIFT1M, for $k=10$, aggregation accounts for less than $3\%$ of total query time. For $k=100$, its share increases from $8.7\%$ at $m=16$ to $24.2\%$ at $m=128$, as aggregation processes the $mk$ candidates returned by the trees.

Besides $k$, the dimension parameter $d$ also impacts the search-to-aggregation ratio. SIFT1M and MSONG both contain approximately one million points and use nearly identical leaf occupancies, while their dimensions are $128$ and $420$, respectively. At $m=128$ and $k=100$, search takes $2.52$~ms on SIFT1M and $7.88$~ms on MSONG, whereas aggregation takes approximately $0.80$~ms on both. Its share of total query time consequently falls from $24.2\%$ to $9.2\%$. Higher dimensionality increases leaf-access and distance-computation costs, whereas aggregation operates only on IDs and distances.

\paragraph{Ablation of selection network}
The DO max heap discussed in Section~\ref{subsec: selection-network} is a viable selector that, like a selection network, retains the exact top-$k$ of a visited leaf. We compare their query costs on SIFT1M for $k=10$ and $k=100$ to evaluate the benefit of network-based selection. Both versions use the same random seed, with all other experimental conditions held constant.

\begin{figure}[htbp]
    \centering
    \includegraphics{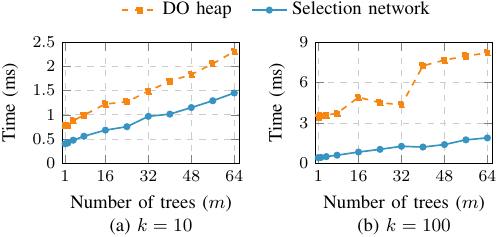}
    \caption{Leaf-selector ablation on SIFT1M.}
    \label{fig:selection-ablation}
\end{figure}

Figure~\ref{fig:selection-ablation} shows query latency for $m=1$ to $64$. The network achieves a $1.53$--$1.92\times$ speedup at $k=10$ and a $3.41$--$7.88\times$ speedup at $k=100$, reducing latency by $34.8$--$47.8\%$ and $70.7$--$87.3\%$, respectively. At $m=64$, latency falls from $2.31$ to $1.45$~ms for $k=10$ and from $8.19$ to $1.90$~ms for $k=100$.

The benefit is concentrated in forest search. At $m=64$, this stage accounts for $0.846$~ms of the $0.855$~ms total reduction at $k=10$ and $6.276$~ms of the $6.298$~ms reduction at $k=100$. Appendix~\ref{app:heap} provides the heap construction, its comparison with Shi's Path Oblivious Heap~\citep{shi2020path}, and the complete ablation results.

\paragraph{Quantitative analysis at high recall} 
Table~\ref{tab: results on high recalls} reports operating points at a target recall of $90\%$, with an $80\%$ target for MS MARCO and for SIFT100M at $k=100$. We select the fastest measured O2N2 configuration meeting the target. For DEEP10M, $m=128$ reaches $90.09\%$ Recall@$10$ in $4.36$~ms. The SIFT100M experiment has no completed baseline. OHNSW exhausted available memory, while OIVF did not finish the indexing process.
\begin{table}[htbp]
    \centering
    \caption{\bb Comparison for $10$-NN and $100$-NN at the specified target recall, where N/A denotes an unavailable result.}
    \label{tab: results on high recalls}
    \begin{threeparttable}
	    % \small
	    \setlength{\tabcolsep}{1.9pt}
        \begin{tabular}{@{} l c c c c c c r @{}}
	    %\begin{tabularx}{\linewidth}{@{}l c c >{\centering\arraybackslash}X >{\raggedleft\arraybackslash}X >{\raggedleft\arraybackslash}X >{\centering\arraybackslash}X >{\raggedleft\arraybackslash}X >{\raggedleft\arraybackslash}X r@{}}
			       \toprule
\multirow{2}{*}{\textbf{Dataset}\tnote{$\ddagger$}} & 
\multirow{2}{*}{{\bb $k$}} & 
\multicolumn{3}{c}{{\bb O2N2}\tnote{$\ast$}} & 
\multicolumn{2}{c}{{\bb Baseline\tnote{$\dagger$}}} & 
\multirow{2}{*}{\textbf{Speedup}} \\
\cmidrule(lr){3-5} \cmidrule(lr){6-7}
& & {\bb $m$} & {\bb Storage} & {\bb Time} & {\bb Storage} & {\bb Time} & \\
\midrule
	        \multirow{2}{*}{\textbf{DEEP10M}}
			        & $10$ & \ensuremath{128} & 256 & %\ensuremath{90.1}\% & 
                    \ensuremath{4.4} & \multirow{2}{*}{1165} & %\ensuremath{89.9}\% & 
                    \ensuremath{333.8} & \ensuremath{76.6}$\times$ \\
			        & $100$ & \ensuremath{208} & 415 & %\ensuremath{91.1}\% & 
                    \ensuremath{6.9} & & %\ensuremath{91.2}\% & 
                    \ensuremath{497.1} & \ensuremath{72.3}$\times$ \\
	        \midrule
		        \multirow{2}{*}{\textbf{SIFT1M}}
				        & $10$ & \ensuremath{56} & 15 & %\ensuremath{90.1}\% & 
	                    \ensuremath{1.3} & \multirow{2}{*}{40} & %\ensuremath{90}\% & 
	                    \ensuremath{161.3} & \ensuremath{125.9}$\times$ \\
				        & $100$ & \ensuremath{96} & 26 & %\ensuremath{90.3}\% & 
	                    \ensuremath{2.5} & & %\ensuremath{89.6}\% & 
	                    \ensuremath{260.5} & \ensuremath{102.9}$\times$ \\
		        \midrule
			        \multirow{2}{*}{\textbf{SIFT100M}\tnote{$\dddagger$}}
					        & $10$ & 88 & 2325 &
		                    \ensuremath{3.5} & \multirow{2}{*}{N/A} & \multirow{2}{*}{N/A} & \multirow{2}{*}{N/A} \\
					        & $100$ & 104 & 2748 &
		                    \ensuremath{4} & & & \\
		        \midrule
		        \multirow{2}{*}{\textbf{MSONG}}
			        & $10$ & \ensuremath{64} & 55 & %\ensuremath{90.4}\% & 
                    \ensuremath{4.3} & \multirow{2}{*}{49} & %\ensuremath{89.7}\% & 
                    \ensuremath{253.1} & \ensuremath{58.4}$\times$ \\
			        & $100$ & \ensuremath{112} & 85 & %\ensuremath{90.7}\% & 
                    \ensuremath{8.1} & & %\ensuremath{93.4}\% & 
                    \ensuremath{322.1} & \ensuremath{39.6}$\times$ \\
	        \midrule
	        \textbf{MS MARCO}
			        & $10$ & 184 & 2540 & %\ensuremath{82.7}\% & 
                    \ensuremath{53.4} & 1526 & %\ensuremath{79.3}\% & 
                    \ensuremath{522.2} & \ensuremath{9.8}$\times$ \\
        \bottomrule
    \end{tabular}
    \begin{tablenotes}[flushleft]
	        \item[$\ast$] Storage and Time denote storage requirement (GB) and latency (ms), and $m$ denotes the number of trees. Storage values are based on the heap-based index measurements. The target recall is $90\%$, except for SIFT100M at $k=100$ and MS MARCO at $k=10$, where it is $80\%$. We report the fastest measured O2N2 configuration meeting the target and the baseline point with recall closest to the target.
	        \item[$\dagger$] We use the fastest baseline OHNSW (Circuit ORAM) here.
	        \item[$\ddagger$] Our machine with 3TB storage can only support up to 200 trees for MS MARCO dataset, so only $k=10$ is stated here.
	        \item[$\dddagger$] SIFT100M baseline results are unavailable because OHNSW exhausts the available storage, while OIVF's index construction was prohibitively slow at this scale and did not finish in our experiments.
    \end{tablenotes}
    \end{threeparttable}
\end{table}

\paragraph{Storage overhead}
As shown in Table~\ref{tab: results on high recalls}, \name{} has clear storage advantage than the baselines for low-dimension datasets (SIFT and DEEP). However, on high-dimensional datasets (MSONG and MS MARCO), the limitation of tree-based methods becomes apparent. To fetch more candidates and achieve higher recall, the number of trees $m$ (and hence the total storage) must increase, causing \name{} to exceed the baselines in storage. In contrast, OHNSW can maintain a fixed storage size: adjusting the query parameter only affects the computation time, not the storage. Our 3 TB storage limit prevents \name{} from reaching $90\%$ Recall@100 on SIFT100M and higher-recall operating points on MS MARCO. Within the feasible parameter regime of MS MARCO, \name{} is still \ensuremath{9.8}$\times$ faster than OHNSW, albeit at a higher storage cost ($2540$~GB vs.\ $1526$~GB).

\subsection{Comparison with existing private ANN systems}
\label{subsec:existing-private-ann}

We compare \name{} with Compass~\cite{zhu2025compass}, Onyx~\cite{rathee2026onyx}, and BiSON~\cite{das2026bison}. Table~\ref{tab:existing-private-ann} places our measurements alongside their published results for $k=10$ and includes Circuit OHNSW. For \name{} on MS MARCO, we report the point with the highest measured recall. The table records the privacy guarantee, recall, and query latency of each reported operating point.

\begin{table}[t]
    \centering
    \setlength{\tabcolsep}{4.8pt}
    \begin{threeparttable}
    \caption{\bb Query latency of private ANN systems for $k=10$.}
    \label{tab:existing-private-ann}
    \begin{tabular*}{\linewidth}{@{\extracolsep{\fill}}l c r r r r@{}}
        \toprule
        \multirow{2}{*}{\textbf{System}\tnote{$\dagger$}} &
        \multirow{2}{*}{\makecell[l]{\bf Privacy}} &
        \multicolumn{2}{c}{\textbf{SIFT1M}} &
        \multicolumn{2}{c}{\textbf{MS MARCO}} \\
        \cmidrule(lr){3-4}\cmidrule(l){5-6}
        & & Recall & Time & Recall & Time \\
        \midrule
        \name{} & Double O & \ensuremath{90.09} & \ensuremath{1.28} & \ensuremath{85.26} & \ensuremath{55.53} \\
        OHNSW & Double O & \ensuremath{90} & \ensuremath{161.3} & \ensuremath{90.24} & \ensuremath{1{,}210.4} \\
        \midrule
        Compass~\cite{zhu2025compass} & (Single) O & $\geq 90$ & $40$ & $\geq 90$ & $470$ \\
        BiSON~\cite{das2026bison} & (Single) O & $90.24$ & $9$ & $91.43$ & $40$ \\
        TEE Compass~\cite{rathee2026onyx} & (Single) O & $\approx 90$ & $42.5$ & $\approx 90$ & $827$ \\
        Onyx~\cite{rathee2026onyx}\tnote{$\ddagger$} & (Single) O & $\approx 90$ & $18.8$ & $\approx 90$ & $67.1$ \\
        \bottomrule
    \end{tabular*}
    \begin{tablenotes}[flushleft]
        \item[$\dagger$] Compass results use the semi-honest fast-network setting (3~Gbps, 1~ms RTT), and BiSON results use the LAN setting (10~Gbps, 0.3~ms RTT). We select these low-latency network settings to limit communication overhead in comparison with our local query execution. For both systems, we report full latency including deferred ORAM eviction, since our query timings also include eviction.
        \item[$\ddagger$] Onyx results use the 1-SSD setting.
    \end{tablenotes}
    \end{threeparttable}
\end{table}

On SIFT1M, the results in Table~\ref{tab:existing-private-ann} show that \name{} is \ensuremath{31.2}$\times$, \ensuremath{7}$\times$, and \ensuremath{14.7}$\times$ faster than Compass, BiSON, and Onyx, respectively. This advantage does not carry over uniformly to the higher-dimensional MS MARCO workload, where the forest requires more trees for high recall. Although \name{} remains faster than Compass and Compass-in-TEE, its highest-recall operating point has latency between BiSON and Onyx and lower recall than both. In addition, note that at approximately $90\%$ recall on MS MARCO, our circuit OHNSW takes only about \ensuremath{1.5}$\times$ as long as Compass-in-TEE while additionally protecting internal memory accesses, demonstrating the competitiveness of our baseline.

The different privacy guarantees help explain these results. Compass and BiSON keep quantized vector hints, position maps, ORAM metadata, and graph caches in trusted client memory~\cite{zhu2025compass,das2026bison}. Candidate node IDs directly index local hints and cached nodes without hiding these accesses. Their reported client memory on MS MARCO is approximately $500$~MB and $827$~MB, respectively. Onyx similarly keeps compressed traversal hints and ORAM client state in TEE memory, with DRAM access patterns explicitly excluded from its threat model~\cite{rathee2026onyx}. These designs benefit from query-dependent local lookups whose access patterns are assumed hidden. In contrast, \name{} protects both external and internal accesses, retaining competitive latency under this additional requirement.

% \rev{The reported configurations also differ in resources. Our \name{} measurements use 32 query workers and an in-memory index, whereas Onyx and Compass-in-TEE use one vCPU, 4~GB of RAM, and one SSD resource unit. Compass uses an 8-vCPU client and a 64-vCPU server, while BiSON uses four ORAM threads. Both client-server systems keep their outsourced ORAM indexes in server RAM. Table~\ref{tab:existing-private-ann} specifies their network settings and timing scope.}

\paragraph{Comparison with Panther}
Panther~\cite{li2025panther} is a PIR-based private ANN search system that protects query privacy and database privacy in a single-server model. Under its reported SIFT1M setting, Panther takes \SI{1.49}{\second}, whereas \name{} meeting $90\%$ Recall@$10$ takes $\ensuremath{1.28}$~ms, a reported-latency ratio of \ensuremath{1{,}163.4}$\times$. Panther provides a cryptographic comparison point alongside the ORAM-based systems above.

\paragraph{Comparison with plaintext search}
To quantify the performance overhead of \name{}, we compare it against two plaintext schemes, HNSWlib and FAISS-IVF, via ann-benchmark~\citep{aumuller2020ann}. At approximately $90\%$ $\text{Recall}@10$ on SIFT1M, HNSWlib and FAISS-IVF take \SI{0.08}{\milli\second} and \SI{0.4}{\milli\second}, respectively. O2N2 is \ensuremath{3.2}$\times$ slower than FAISS-IVF and \ensuremath{16}$\times$ slower than HNSWlib. We argue that this overhead remains acceptable for remote services, where the round-trip latency constitute the major cost of ANN search.

\section{Conclusion and future work}
We presented \name{}, a doubly oblivious ANN system for TEEs. By aligning RR~\kdt{} traversals with tree-based ORAM, \name{} reduces the overhead per block access from $\bigo(\log^2 N)$ for generic recursive doubly oblivious storage to $\bigo(\log N)$. Experiments across datasets and query parameters show that \name{} achieves \ensuremath{9.8}--\ensuremath{125.9}$\times$ speedup over doubly oblivious baselines at the evaluated high-recall operating points.

Nevertheless, \name{} inherits limitations from its tree-based design. While it is more space- and time-efficient on low-dimensional datasets, achieving high recall on high-dimensional data requires more trees and substantially increases storage costs. Improving this space--latency tradeoff, for example through hybrid index structures, is an interesting direction for future work.

\section*{AI assistance disclosure}
We used GPT-5.6 and GPT-6 to optimize the implementation, prepare experiment scripts, process benchmark logs, and review the proofs. GPT and Grok also assisted with checking grammar and revising the manuscript. The authors take responsibility for the implementation, analysis, and conclusions.

\bibliographystyle{IEEEtran}
\bibliography{ref}

@inproceedings{sinha2014lsh,
  author    = {Sinha, Kaushik},
  booktitle = {International Conference on Machine Learning and
               Applications},
  pages     = {41--46},
  title     = {LSH vs randomized partition trees: Which one to use
               for nearest neighbor search?},
  year      = {2014}
}

@article{aumuller2020ann,
  title   = {ANN-Benchmarks: A benchmarking tool for approximate nearest neighbor algorithms},
  author  = {Aum{\"u}ller, Martin and Bernhardsson, Erik and Faithfull, Alexander},
  journal = {Information Systems},
  volume  = {87},
  pages   = {101374},
  year    = {2020}
}

@article{bajaj2016msmarco,
  title   = {{MS MARCO}: A Human Generated Machine Reading Comprehension Dataset},
  author  = {Bajaj, Payal and Campos, Daniel and Craswell, Nick and Deng, Li and Gao, Jianfeng and Liu, Xiaodong and Majumder, Rangan and McNamara, Andrew and Mitra, Bhaskar and Nguyen, Tri and Rosenberg, Mir and Song, Xia and Stoica, Alina and Tiwary, Saurabh and Wang, Tong},
  journal = {arXiv preprint arXiv:1611.09268},
  year    = {2016}
}

@inproceedings{bertin2011million,
  title     = {The Million Song Dataset},
  author    = {Bertin-Mahieux, Thierry and Ellis, Daniel P. W. and Whitman, Brian and Lamere, Paul},
  booktitle = {Proceedings of {ISMIR}},
  pages     = {591--596},
  year      = {2011}
}

@inproceedings{lin2021pyserini,
  title     = {Pyserini: A Python Toolkit for Reproducible Information Retrieval Research with Sparse and Dense Representations},
  author    = {Lin, Jimmy and Ma, Xueguang and Lin, Sheng-Chieh and Yang, Jheng-Hong and Pradeep, Ronak and Nogueira, Rodrigo},
  booktitle = {Proceedings of {SIGIR}},
  pages     = {2356--2362},
  year      = {2021}
}

@inproceedings{aumuller2019role,
  title        = {The role of local intrinsic dimensionality in benchmarking nearest neighbor search},
  author       = {Aum{\"u}ller, Martin and Ceccarello, Matteo},
  booktitle    = {Proceedings of the International Conference on Similarity Search and Applications},
  pages        = {113--127},
  year         = {2019},
  organization = {Springer}
}

@inproceedings{cong2025revisiting,
  author    = {Cong, Kelong and Geelen, Robin and Kang, Jiayi and Park, Jeongeun},
  title     = {Revisiting Oblivious {Top-k} Selection with Applications to Secure {k-NN} Classification},
  booktitle = {Selected Areas in Cryptography -- SAC 2024},
  series    = {Lecture Notes in Computer Science},
  volume    = {15516},
  pages     = {3--25},
  publisher = {Springer},
  year      = {2025},
  doi       = {10.1007/978-3-031-82852-2_1},
  url       = {https://eprint.iacr.org/2023/852}
}

@inproceedings{batcher1968sorting,
  author    = {Batcher, Kenneth E},
  booktitle = {Proceedings of the Spring Joint Computer Conference},
  pages     = {307--314},
  title     = {Sorting networks and their applications},
  year      = {1968}
}

@article{bentley1975multidimensional,
  author  = {Bentley, Jon Louis},
  journal = {Communications of the {ACM}},
  number  = {9},
  pages   = {509--517},
  title   = {Multidimensional binary search trees used for
             associative searching},
  volume  = {18},
  year    = {1975}
}

@inproceedings{cash2015leakage,
  title     = {Leakage-abuse attacks against searchable encryption},
  author    = {Cash, David and Grubbs, Paul and Perry, Jason and Ristenpart, Thomas},
  booktitle = {Proceedings of {CCS}},
  pages     = {668--679},
  year      = {2015}
}

@article{chamani2023graphos,
  author  = {Javad Ghareh Chamani and Ioannis Demertzis and
             Dimitrios Papadopoulos and Charalampos Papamanthou and
             Rasool Jalili},
  journal = {Proceedings of the {VLDB} Endowment},
  number  = {13},
  pages   = {4324--4338},
  title   = {GraphOS: Towards oblivious graph processing},
  volume  = {16},
  year    = {2023}
}

@inproceedings{chen2020sanns,
  title     = {{SANNS}: Scaling up secure approximate {k-nearest} neighbors search},
  author    = {Chen, Hao and Chillotti, Ilaria and Dong, Yihe and Poburinnaya, Oxana and Razenshteyn, Ilya and Riazi, M Sadegh},
  booktitle = {Proceedings of {USENIX} Security},
  pages     = {2111--2128},
  year      = {2020}
}

@inproceedings{chuang2026tee,
  author    = {Chuang, J and Seto, A and Berrios, N and
               van Schaik, S and Garman, C and Genkin, D},
  booktitle = {Proceedings of {SP}},
  pages     = {1894--1912},
  title     = {Tee. fail: Breaking trusted execution environments
               via ddr5 memory bus interposition},
  year      = {2026}
}

@article{DBLP:journals/compsec/RachidRM20,
  author  = {Maan Haj Rachid and
             Ryan D. Riley and
             Qutaibah M. Malluhi},
  title   = {Enclave-based oblivious {RAM} using Intel's {SGX}},
  journal = {Computers \& Security},
  volume  = {91},
  pages   = {101711},
  year    = {2020}
}

@article{feng2024survey,
  title   = {Survey of research on confidential computing},
  author  = {Feng, Dengguo and Qin, Yu and Feng, Wei and Li, Wei and Shang, Ketong and Ma, Hongzhan},
  journal = {IET Communications},
  volume  = {18},
  number  = {9},
  pages   = {535--556},
  year    = {2024}
}

@inproceedings{goldreich1987towards,
  author    = {Goldreich, Oded},
  booktitle = {Proceedings of {STOC}},
  pages     = {182--194},
  title     = {Towards a theory of software protection and
               simulation by oblivious RAMs},
  year      = {1987}
}

@inproceedings{grubbs2016breaking,
  title     = {Breaking web applications built on top of encrypted data},
  author    = {Grubbs, Paul and McPherson, Richard and Naveed, Muhammad and Ristenpart, Thomas and Shmatikov, Vitaly},
  booktitle = {Proceedings of {CCS}},
  pages     = {1353--1364},
  year      = {2016}
}

@inproceedings{henzinger2023private,
  title     = {Private web search with Tiptoe},
  author    = {Henzinger, Alexandra and Dauterman, Emma and Corrigan-Gibbs, Henry and Zeldovich, Nickolai},
  booktitle = {Proceedings of {SOSP}},
  pages     = {396--416},
  year      = {2023}
}

@techreport{intel-tdx-whitepaper-2022,
  author      = {{Intel Corporation}},
  title       = {Intel® trust domain extensions},
  year        = {2022},
  month       = {February},
  institution = {Intel Corporation},
  url         = {https://cdrdv2-public.intel.com/690419/TDX-Whitepaper-February2022.pdf},
  note        = {White Paper}
}

@inproceedings{islam2012access,
  author    = {Islam, Mohammad Saiful and Kuzu, Mehmet and
               Kantarcioglu, Murat},
  booktitle = {Proceedings of {NDSS}},
  pages     = {12},
  title     = {Access pattern disclosure on searchable encryption:
               Ramification, attack and mitigation},
  volume    = {20},
  year      = {2012}
}

@article{jayaram2019diskann,
  title   = {{DiskANN}: Fast accurate billion-point nearest neighbor search on a single node},
  author  = {Jayaram Subramanya, Suhas and Devvrit, Fnu and Simhadri, Harsha Vardhan and Krishnawamy, Ravishankar and Kadekodi, Rohan},
  journal = {Advances in Neural Information Processing Systems},
  volume  = {32},
  year    = {2019}
}

@inproceedings{simhadri2022results,
  title     = {Results of the {NeurIPS'21} Challenge on Billion-Scale Approximate Nearest Neighbor Search},
  author    = {Simhadri, Harsha Vardhan and Williams, George and Aum{\"u}ller, Martin and Douze, Matthijs and Babenko, Artem and Baranchuk, Dmitry and Chen, Qi and Hosseini, Lucas and Krishnaswamy, Ravishankar and Srinivasa, Gopal and Subramanya, Suhas Jayaram and Wang, Jingdong},
  booktitle = {Proceedings of the NeurIPS 2021 Competitions and Demonstrations Track},
  volume    = {176},
  pages     = {177--189},
  year      = {2022}
}

@article{jegou2010product,
  title   = {Product quantization for nearest neighbor search},
  author  = {Jegou, Herve and Douze, Matthijs and Schmid, Cordelia},
  journal = {IEEE Transactions on Pattern Analysis and Machine Intelligence},
  volume  = {33},
  number  = {1},
  pages   = {117--128},
  year    = {2010}
}

@article{johnson2019billion,
  title   = {Billion-scale similarity search with GPUs},
  author  = {Johnson, Jeff and Douze, Matthijs and J{\'e}gou, Herv{\'e}},
  journal = {{IEEE} Transactions on Big Data},
  volume  = {7},
  number  = {3},
  pages   = {535--547},
  year    = {2019}
}

@techreport{kaplan2016amd,
  author      = {Kaplan, David and Powell, Jeremy and Woller, Tom},
  title       = {{AMD Memory Encryption}},
  institution = {Advanced Micro Devices, Inc.},
  year        = {2016},
  month       = {May},
  note        = {White Paper}
}

@article{lewis2020retrieval,
  title   = {Retrieval-augmented generation for knowledge-intensive nlp tasks},
  author  = {Lewis, Patrick and Perez, Ethan and Piktus, Aleksandra and Petroni, Fabio and Karpukhin, Vladimir and Goyal, Naman and K{\"u}ttler, Heinrich and Lewis, Mike and Yih, Wen-tau and Rockt{\"a}schel, Tim and others},
  journal = {Advances in Neural Information Processing Systems},
  volume  = {33},
  pages   = {9459--9474},
  year    = {2020}
}

@inproceedings{li2025panther,
  title     = {{Panther}: Private approximate nearest neighbor search in the single server setting},
  author    = {Li, Jingyu and Huang, Zhicong and Zhang, Min and Hong, Cheng and Liu, Jian and Wei, Tao and Chen, Wenguang},
  booktitle = {Proceedings of {CCS}},
  pages     = {365--379},
  year      = {2025}
}

@inproceedings{mckeen2013innovative,
  author    = {Frank McKeen and
               Ilya Alexandrovich and
               Alex Berenzon and
               Carlos V. Rozas and
               Hisham Shafi and
               Vedvyas Shanbhogue and
               Uday R. Savagaonkar},
  title     = {Innovative instructions and software model for isolated execution},
  booktitle = {{HASP@ISCA}},
  pages     = {10},
  year      = {2013}
}

@article{mikolov2013distributed,
  title   = {Distributed representations of words and phrases and their compositionality},
  author  = {Mikolov, Tomas and Sutskever, Ilya and Chen, Kai and Corrado, Greg S and Dean, Jeff},
  journal = {Advances in Neural Information Processing Systems},
  volume  = {26},
  year    = {2013}
}

@inproceedings{mishra2018oblix,
  author    = {Pratyush Mishra and
               Rishabh Poddar and
               Jerry Chen and
               Alessandro Chiesa and
               Raluca Ada Popa},
  title     = {{Oblix}: An efficient oblivious search index},
  booktitle = {Proceedings of {SP}},
  pages     = {279--296},
  year      = {2018}
}

@inproceedings{ram2019revisiting,
  author    = {Ram, Parikshit and Sinha, Kaushik},
  booktitle = {Proceedings of {SIGKDD}},
  pages     = {1378--1388},
  title     = {Revisiting kd-tree for nearest neighbor search},
  year      = {2019}
}

@inproceedings{ren2015constants,
  author    = {Ren, Ling and Fletcher, Christopher and Kwon, Albert and
               Stefanov, Emil and Shi, Elaine and Van Dijk, Marten and
               Devadas, Srinivas},
  booktitle = {Proceedings of {USENIX} Security},
  pages     = {415--430},
  title     = {Constants count: Practical improvements to oblivious
               {RAM}},
  year      = {2015}
}

@article{sardar2023confidential,
  title   = {Confidential computing and related technologies: a critical review},
  author  = {Sardar, Muhammad Usama and Fetzer, Christof},
  journal = {Cybersecurity},
  volume  = {6},
  number  = {1},
  pages   = {10},
  year    = {2023}
}

@inproceedings{servan2022private,
  title     = {Private approximate nearest neighbor search with sublinear communication},
  author    = {Servan-Schreiber, Sacha and Langowski, Simon and Devadas, Srinivas},
  booktitle = {Proceedings of {SP}},
  pages     = {911--929},
  year      = {2022}
}

@inproceedings{shaul2020secure,
  title     = {Secure $k$-ish nearest neighbors classifier},
  author    = {Shaul, Hayim and Feldman, Dan and Rus, Daniela},
  booktitle = {Proceedings on PETS},
  volume    = {3},
  pages     = {42--61},
  year      = {2020}
}

@inproceedings{shi2020path,
  author    = {Elaine Shi},
  booktitle = {Proceedings of {SP}},
  pages     = {842--858},
  title     = {Path oblivious heap: Optimal and practical oblivious
               priority queue},
  year      = {2020}
}

@inproceedings{song2023secure,
  title     = {Secure approximate nearest neighbor search with locality-sensitive hashing},
  author    = {Song, Shang and Liu, Lin and Chen, Rongmao and Peng, Wei and Wang, Yi},
  booktitle = {Proceedings of {ESORICS}},
  pages     = {411--430},
  year      = {2023}
}

@article{stefanov2018path,
  author  = {Emil Stefanov and Marten van Dijk and Elaine Shi and
             T.{-}H. Hubert Chan and Christopher W. Fletcher and
             Ling Ren and Xiangyao Yu and Srinivas Devadas},
  journal = {Journal of {ACM}},
  number  = {4},
  pages   = {18:1--18:26},
  title   = {Path {ORAM:} An extremely simple oblivious {RAM}
             protocol},
  volume  = {65},
  year    = {2018}
}

@inproceedings{tinoco2023enigmap,
  author    = {Tinoco, Afonso and Gao, Sixiang and Shi, Elaine},
  booktitle = {Proceedings of {USENIX} Security},
  pages     = {4033--4050},
  title     = {{EnigMap}: {External-Memory} oblivious map
               for secure enclaves},
  year      = {2023}
}

@inproceedings{van2017telling,
  author    = {Van Bulck, Jo and Weichbrodt, Nico and Kapitza, R{\"u}diger and Piessens, Frank and Strackx, Raoul},
  booktitle = {Proceedings of {USENIX} Security},
  pages     = {1041--1056},
  title     = {Telling your secrets without page faults: Stealthy
               page {Table-Based} attacks on enclaved
               execution},
  year      = {2017}
}

@inproceedings{van2018foreshadow,
  title     = {Foreshadow: Extracting the keys to the intel {SGX} kingdom with transient {Out-of-Order} execution},
  author    = {Van Bulck, Jo and Minkin, Marina and Weisse, Ofir and Genkin, Daniel and Kasikci, Baris and Piessens, Frank and Silberstein, Mark and Wenisch, Thomas F and Yarom, Yuval and Strackx, Raoul},
  booktitle = {Proceedings of {USENIX} Security},
  pages     = {991--1008},
  year      = {2018}
}

@misc{van2020sgaxe,
  author = {Van Schaik, Stephan and Kwong, Andrew and
            Genkin, Daniel and Yarom, Yuval},
  title  = {SGAxe: How SGX fails in practice},
  year   = {2020}
}

@inproceedings{wang2014oblivious,
  author    = {Xiao Shaun Wang and Kartik Nayak and Chang Liu and
               T.{-}H. Hubert Chan and Elaine Shi and Emil Stefanov and
               Yan Huang},
  booktitle = {Proceedings of {CCS}},
  pages     = {215--226},
  title     = {Oblivious data structures},
  year      = {2014}
}

@inproceedings{wang2015circuit,
  author    = {Wang, Xiao and Chan, Hubert and Shi, Elaine},
  booktitle = {Proceedings of {CCS}},
  pages     = {850--861},
  title     = {Circuit {ORAM}: On tightness of the
               Goldreich-Ostrovsky lower bound},
  year      = {2015}
}

@inproceedings{xu2015controlled,
  author    = {Xu, Yuanzhong and Cui, Weidong and Peinado, Marcus},
  booktitle = {Proceedings of {SP}},
  pages     = {640--656},
  title     = {Controlled-channel attacks: Deterministic side
               channels for untrusted operating systems},
  year      = {2015}
}

@inproceedings{zhang2016all,
  author    = {Zhang, Yupeng and Katz, Jonathan and
               Papamanthou, Charalampos},
  booktitle = {Proceedings of {USENIX} Security},
  pages     = {707--720},
  title     = {All your queries are belong to us: the power of
               {File-Injection} attacks on searchable
               encryption},
  year      = {2016}
}

@inproceedings{zheng2025h2o2ram,
  author    = {Zheng, Leqian and Zhang, Zheng and Dong, Wentao and
               Zhang, Yao and Wu, Ye and Wang, Cong},
  booktitle = {Proceedings of {USENIX} Security},
  pages     = {8501--8520},
  title     = {{H2O2RAM}: A {High-Performance}
               hierarchical doubly oblivious {RAM}},
  year      = {2025}
}

@inproceedings{zhou2024pacmann,
  author    = {Zhou, Mingxun and Shi, Elaine and Fanti, Giulia},
  title     = {{Pacmann}: Efficient private approximate nearest neighbor search},
  booktitle = {Proceedings of {ICLR}},
  year      = {2025}
}

@article{zhu2024confidential,
  title   = {Confidential Computing on NVIDIA Hopper GPUs: A Performance Benchmark Study},
  author  = {Zhu, Jianwei and Yin, Hang and Deng, Peng and Almeida, Aline and Zhou, Shunfan},
  journal = {arXiv preprint arXiv:2409.03992},
  year    = {2024}
}

@inproceedings{das2026bison,
  title     = {{BiSON}: Billion-Scale Oblivious Nearest-Neighbor Search in Milliseconds},
  author    = {Das, Sankha and Ravi, Rohan and Chandran, Nishanth and Gupta, Divya},
  booktitle = {2026 IEEE 11th European Symposium on Security and Privacy (EuroS\&P)},
  pages     = {160--172},
  publisher = {IEEE},
  year      = {2026},
  doi       = {10.1109/EuroSP68448.2026.00021},
  url       = {https://eprint.iacr.org/2026/1343}
}

@article{rathee2026onyx,
  title   = {{Onyx}: Cost-Efficient Disk-Oblivious {ANN} Search},
  author  = {Rathee, Deevashwer and Watson, Jean-Luc and Zhao, Zirui Neil and Suh, G. Edward and Popa, Raluca Ada},
  journal = {arXiv preprint arXiv:2604.20401},
  year    = {2026}
}

@inproceedings{zhu2025compass,
  title     = {{Compass}: Encrypted semantic search with high accuracy},
  author    = {Zhu, Jinhao and Patel, Liana and Zaharia, Matei and Popa, Raluca Ada},
  booktitle = {Proceedings of {OSDI}},
  pages     = {915--938},
  year      = {2025}
}

@inproceedings{zuber2021efficient,
  author    = {Zuber, Martin and Sirdey, Renaud},
  title     = {Efficient homomorphic evaluation of k-NN classifiers},
  booktitle = {Proceedings of {PETS}},
  volume    = {2021},
  number    = {2},
  pages     = {111--129},
  year      = {2021}
}

\clearpage
\appendix

% \noindent {\bf\LARGE Technical Appendix for \name{}}

\section{Heap alternative and supplementary ablation}
\label{app:heap}

The main construction uses an oblivious selection network to obtain the top-$k$ candidates from each visited leaf. This appendix describes the doubly oblivious (DO) heap used as the alternative selector in the ablation. We first report the heap-operation comparison with Shi's Path OHeap~\cite{shi2020path}, then describe the heap and compare its query latency with the network version.

\subsection{Comparison with Path OHeap}
The DO heap supports fixed-capacity top-$k$ updates: a new candidate replaces the current maximum only when its distance is smaller. We compare it with the official Path OHeap implementation\footnote{\url{https://github.com/obliviousram/PathOHeap}} by filling both structures with $k$ blocks of the same size and performing the same number of extract-maximum/insert update pairs. Table~\ref{tab:heap-small-comparison} retains the measured operation times. Relative to the fastest reported Path OHeap configuration at each capacity, our heap is $2.79\times$, $1.77\times$, and $1.17\times$ faster for $k=10$, $100$, and $1000$, respectively. Thus this implementation is competitive at small capacities, although the measured advantage narrows as $k$ grows.

\begin{table}[htbp]
    \centering
    \caption{Heap update latency ($\mu$s per extract/insert pair). $Z$ is the Path OHeap bucket size.}
    \label{tab:heap-small-comparison}
    \begin{tabular}{@{}rrrrrr@{}}
        \toprule
        $k$ & $Z=1$ & $Z=2$ & $Z=4$ & DO heap & Speedup \\
        \midrule
        10 & \textbf{3.902} & 4.641 & 4.572 & 1.400 & $2.79\times$ \\
        100 & 6.211 & \textbf{6.181} & 7.123 & 3.493 & $1.77\times$ \\
        1000 & 10.464 & \textbf{8.553} & 10.116 & 7.284 & $1.17\times$ \\
        \bottomrule
    \end{tabular}
\end{table}

\subsection{DO heap construction}
The heap is a pointer-based binary tree stored in a DO Circuit ORAM~\cite{wang2015circuit}. Its capacity $c$ is public. Set $L=\lceil\log_2(c+1)\rceil$ and pad the tree to $2^L-1$ nodes. The first $c$ nodes are active slots; the remaining nodes are inactive padding. A node stores an item--distance pair $(x,v)$ and pointers to its children. Each pointer contains both a logical identifier and its ORAM position. Child positions are embedded in parent blocks following the pointer-based access technique of Wang et al.~\cite{wang2014oblivious}. The stash is scanned on a fixed schedule, and local comparisons and record moves use masked operations.

For top-$k$ selection, set $c=k$. Initially, the active slots contain dummy candidates with value $+\infty$; inactive slots have value $-\infty$. For each of the public number $n_0$ of leaf slots, compute its distance and execute a fixed-capacity update. A valid candidate replaces the root exactly when it is better than the current maximum. A padded leaf slot is represented by an invalid candidate with value $+\infty$. After the scan, exactly $k$ extractions return the retained candidates. The forest uses the same cross-tree duplicate removal and final selection as the network version.

The basic interface consists of $\mathsf{Insert}(x,v)$, $\mathsf{ExtractMax}()$, and a fixed-height $\mathsf{OSift}$ operation. Insert conditionally replaces the root and always runs OSift. ExtractMax saves the root record, replaces it with $(\dummy,-\infty)$, runs OSift, and returns the saved record. After all inputs have been offered, this yields the best $k$ leaf candidates; reversing the extraction order produces ascending distances.

\SetKwFunction{OInsert}{Insert}
\SetKwFunction{OExtractMax}{ExtractMax}
\SetKwFunction{OExtractAll}{ExtractAll}
\SetKwFunction{OHeapify}{Heapify}
\SetKwFunction{OSift}{OSift}
\begin{algorithm}[htbp]
    \caption{Doubly oblivious max heap}
    \label{alg: omaxheap}
    \KwIn{Doubly oblivious max heap $\mathcal{H}=(\mathcal{T},c,L,\root)$}
    \Fn{\OInsert{$x$, $v$}}{
        $b_{\root}=(x_{\root}, v_{\root}, \lptr, \rptr)\gets \mathcal{T}.\mathsf{ReadRm}(\root)$\;
        \If{$v$ is smaller than $v_{\root}$ under the public tie-breaking rule}{
            $b_{\root}\gets (x,v,\lptr,\rptr)$\;
        }
        Assign new $\pos$ to $\root$ and call $\mathcal{T}.\mathsf{Write}(\root,b_{\root})$\;
        Call $\mathcal{H}.\OSift(\root)$\;
    }
    \Fn{\OExtractMax{}}{
        $b_{\root}=(x_{\root}, v_{\root}, \lptr, \rptr)\gets \mathcal{T}.\mathsf{ReadRm}(\root)$\;
        $b_{\root}\gets (\dummy, -\infty, \lptr, \rptr)$\;
        Assign new $\pos$ to $\root$ and call $\mathcal{T}.\mathsf{Write}(\root,b_{\root})$\;
        Call $\mathcal{H}.\OSift(\root)$\;
        \KwRet{$(x_{\root}, v_{\root})$}
    }
    \Fn{\OSift{$\ptr$}}{
        \For{$i\gets 1$ \KwTo $L-1$}{
            $b=(x,v,\lptr,\rptr)\gets \mathcal{T}.\mathsf{ReadRm}(\ptr)$\;
            $b_l=(x_l,v_l,\ldots)\gets \mathcal{T}.\mathsf{ReadRm}(\lptr)$\;
            $b_r=(x_r,v_r,\ldots)\gets \mathcal{T}.\mathsf{ReadRm}(\rptr)$\;
            Assign new $\pos$ values to $\ptr$, $\lptr$, and $\rptr$\;
            $j\gets\arg\max\{v,v_l,v_r\}$\;
            \If{$j=\mathsf{left}$}{
                Swap the $(x,v)$-entries in $b$ and $b_l$\;
                $\ptr_{\mathsf{next}}\gets\lptr$\;
            }\ElseIf{$j=\mathsf{right}$}{
                Swap the $(x,v)$-entries in $b$ and $b_r$\;
                $\ptr_{\mathsf{next}}\gets\rptr$\;
            }\Else{
                $\ptr_{\mathsf{next}}\gets$ randomly chosen from $\{\lptr,\rptr\}$\;
            }
            Write $b$, $b_l$, and $b_r$ back using their updated pointers\;
            $\ptr\gets\ptr_{\mathsf{next}}$\;
        }
    }
\end{algorithm}

Algorithm~\ref{alg: omaxheap} gives the three operations. OSift executes $L-1$ rounds, each with three $\mathsf{ReadRm}$ calls followed by three $\mathsf{Write}$ calls. Conditional updates and swaps denote masked operations on fixed local records. If no swap is needed, the algorithm continues into a randomly chosen child and completes the same number of rounds.

Position reassignment uses a future-position convention. For a child revisited in the next round, its parent records a reserved final position, while the child is temporarily written at an independent fresh position retained by the traversal. The next round reads from this temporary position and writes the current node at its reserved position, without resampling that position. Other children, including both children in the final round, are written directly to the positions recorded by their parent. Only the current read pointer and reserved output position need to be carried between rounds. The heap's root pointer is updated to the root's final output position.

For $c,k\ge2$, an ORAM access to a $B$-word block costs $\bigo(BZ\log c)$ for bucket size $Z$. OSift makes $\bigo(\log c)$ accesses, so Insert and ExtractMax each cost $\bigo(BZ\log^2 c)$. A leaf scan and final extraction therefore cost $\bigo((n_0+k)BZ\log^2 k)$. In the compact implementation, $x$ is a point identifier, while the vector itself remains in the leaf buffer. Root handling is fused with the update and the workspace is reused between queries. These choices are enabled in the heap endpoint of the ablation.

\subsection{Correctness and obliviousness}
\begin{claim}
If the two subtrees below the root satisfy the max-heap property, a complete OSift restores that property throughout the heap without changing its multiset of records.
\end{claim}
\begin{proof}
At the beginning of each real sift step, only an edge from the current node to one of its children can violate the heap property. Moving the largest of the three values to the current node repairs both outgoing edges. Only the selected child's outgoing edges can then require repair. If the current node already holds the maximum, the entire remaining subtree is ordered and the dummy continuation leaves it unchanged. The fixed $L-1$ rounds reach the bottom level. Masked swaps preserve the multiset. For a revisited child, the traversal uses its temporary read position and then writes it at the reserved position recorded by its parent. All other child pointers name their immediate write positions. Thus every parent--child link is valid when OSift completes.
\end{proof}

Insert leaves the retained multiset unchanged or replaces its maximum with a better candidate. Induction over the leaf scan shows that the active slots retain the best $k$ records seen so far, counting initial $+\infty$ dummies when necessary. ExtractMax returns the current maximum and replaces it with $-\infty$, so exactly $k$ extractions enumerate the retained records. The tie convention can be fixed by ordering distance--ID pairs; the measured older heap endpoint used distance-only comparisons.

\begin{claim}
Assuming the DO Circuit ORAM interface and fixed-schedule local operations, the fixed-capacity heap updates and extractions are doubly oblivious.
\end{claim}
\begin{proof}
For public height $L$, OSift issues exactly $6(L-1)$ ORAM calls; Insert and ExtractMax each add one $\mathsf{ReadRm}$ and one $\mathsf{Write}$. Position samples use fixed-size local buffers and a fixed schedule. Candidate values affect only masked record contents and logical pointers, not local addresses or operation types. Adaptive doubly ORAM security hides the resulting logical access sequence, and fixed local scans hide internal record selection. Sequential composition over the public number of updates and extractions establishes the claim.
\end{proof}

\subsection{Heap versus selection network}
The query-level ablation uses SIFT1M, $k\in\{10,100\}$, seed~1, $1000$ queries, and $32$ query workers. Each endpoint builds $64$ trees with the same forest seed and varies the queried prefix. Both use FP16 leaf payloads, disable initial random candidates, and retain the same tree-search and final aggregation routines. The heap endpoint enables compact distance--ID records, fused root handling, and workspace reuse. The network endpoint uses our blockwise bitonic selector, replacing the per-leaf heap by block sorting and truncated merges over the complete leaf.

Table~\ref{tab:app-selector} gives the $11$ measured prefixes for each value of $k$. The network achieves $1.53$--$1.92\times$ speedup at $k=10$ and $3.41$--$7.88\times$ at $k=100$. Recall agrees at the printed precision except at $k=10$, $m=64$, where it is $90.00\%$ for the heap and $90.01\%$ for the network. Distance-sum ratio and RDE agree at all reported points for both values of $k$. The network evaluates the full leaf candidate set and avoids the heap's per-candidate ORAM updates and final extraction phase.

\begin{table}[htbp]
    \centering
    \caption{Paired SIFT1M query latency for $k=10$ and $k=100$. Times are milliseconds.}
    \label{tab:app-selector}
    \setlength{\tabcolsep}{2.8pt}
    \begin{tabular}{rrrrrrr}
        \toprule
        & \multicolumn{3}{c}{$k=10$} & \multicolumn{3}{c}{$k=100$}\\
        \cmidrule(lr){2-4}\cmidrule(lr){5-7}
        $m$ & DO heap & Network & Speedup & DO heap & Network & Speedup\\
        \midrule
        $1$ & $0.780$ & $0.407$ & $1.92$ & $3.434$ & $0.436$ & $7.88$ \\
        $2$ & $0.786$ & $0.421$ & $1.87$ & $3.535$ & $0.450$ & $7.86$ \\
        $4$ & $0.881$ & $0.479$ & $1.84$ & $3.583$ & $0.513$ & $6.99$ \\
        $8$ & $0.995$ & $0.561$ & $1.77$ & $3.720$ & $0.617$ & $6.03$ \\
        $16$ & $1.225$ & $0.687$ & $1.78$ & $4.906$ & $0.849$ & $5.78$ \\
        $24$ & $1.268$ & $0.757$ & $1.68$ & $4.508$ & $1.049$ & $4.30$ \\
        $32$ & $1.489$ & $0.971$ & $1.53$ & $4.343$ & $1.274$ & $3.41$ \\
        $40$ & $1.695$ & $1.016$ & $1.67$ & $7.236$ & $1.220$ & $5.93$ \\
        $48$ & $1.831$ & $1.151$ & $1.59$ & $7.658$ & $1.403$ & $5.46$ \\
        $56$ & $2.051$ & $1.292$ & $1.59$ & $7.962$ & $1.753$ & $4.54$ \\
        $64$ & $2.306$ & $1.451$ & $1.59$ & $8.194$ & $1.896$ & $4.32$ \\
        \bottomrule
    \end{tabular}
\end{table}

These two experiments measure different workloads: Table~\ref{tab:heap-small-comparison} compares heap operations, whereas Table~\ref{tab:app-selector} measures full forest queries with alternative local selectors. Together they explain the design choice: the DO heap is effective among the tested small-capacity heap implementations, while the fixed leaf-selection workload is served more efficiently by the network.

\end{document}